\documentclass[preprint,amsmath,floatfix,prb,aps]{revtex4-1}

\usepackage{xcolor}
\usepackage{amsmath}
\usepackage{pifont}   
\usepackage{graphicx} 
\usepackage{dcolumn}  
\usepackage{bm}       
\usepackage{amsfonts} 
\usepackage{amssymb}  
\usepackage{multirow} 
\usepackage{romannum}
\usepackage{siunitx}
\usepackage{mathtools}

    \usepackage{braket}
    \usepackage{nicefrac}
    \usepackage{bm}
    \usepackage{physics}

    \renewcommand{\v}[1]{\bm{\mathrm{#1}}}
    \newcommand{\m}[1]{\bm{\mathsf{#1}}}

\usepackage[english]{babel}
\usepackage[autostyle, english = american]{csquotes}
\MakeOuterQuote{"}

\begin{document}

\title{Valley control by linearly polarized laser pulses}

\author{S. Sharma$^1$}
\author{P. Elliott$^1$}
\author{S. Shallcross$^1$}
\email{shallcross@mbi-berlin.de}
\affiliation{1 Max-Born-Institute for non-linear optics, Max-Born Strasse 2A, 12489 Berlin, Germany}

\date{\today}

\begin{abstract}
Underpinning the field of "valleytronics" is the coupling of the helicity of circularly polarized light to the valley degree of freedom, and this remains the only known lightform to exhibit this remarkable effect. Here we show that on femtosecond time scales valley coupling is a much more general effect. We find that two time separated linearly polarized pulses allow almost complete control over valley excitation, with the pulse time difference and polarization vectors emerging as key parameters for valley control. In contrast to the Berry curvature that underpins the effect for circularly polarized light, we demonstrate that a different phase structure drives this effect, with excitations during each linear pulse acquiring a valley discriminating phase involving the polarisation angle of linear light. Unimportant in a single linear pulse, for pairs of pulses these can constructively and destructively interfere. Employing state-of-the-art time dependent density function theory, we show that the effect is robust to the complexities of charge dynamics in a real material with the example of a transitional metal dichalcogenide.
\end{abstract}


\maketitle

\section{Introduction}

A prominent role in the electronic properties of many two dimensional materials is played by the valley degree of freedom\cite{SSZA14,mak_lightvalley_2018,chen_chiral_2020}. Controlling this degree of freedom promises a novel valley based electronics ("valleytronics"), and thus over the last decade intense research activity has centred on the search for valley control, both by material modification via complex deformations\cite{settnes_graphene_2016,gupta_straintronics_2019,zhao_ultrafast_2022,yang_chiral_2019} as well as by external fields such as light
\cite{xiao_coupled_2012,mak_control_2012,xiao_nonlinear_2015,
langer_lightwave_2018,berghauser_inverted_2018,ishii_optical_2019}. This latter route is exemplified by the celebrated spin-valley coupling of certain strong spin-orbit transition metal dichalcogenides, in which the helicity of circularly polarised light determines the spin character and valley at which charge is excited. This physics arises from the fact that valleys are endowed with Berry curvature of opposite sign at conjugate valleys\cite{cho_experimental_2018}, from which follows a valley selection rule for the dipole matrix elements with circular light\cite{xiao_coupled_2012}. Such a selection rule would, however, appear to rule out the possibility of valley control by any other light waveform, e.g. linearly polarized pulses, sharply circumscribing the application to this field of the rich control over light waveforms that modern lasers provide.

What we wish to show here is that, contrary to this common belief, ultrafast pulses of linearly polarized light can provide almost complete valley control; in materials with spin-orbit split valley bands this will lead to spin-valley coupling.
We show that overlapping and temporally separated pulses of orthogonal linear polarisation excite either at the K or K$^\ast$ valley, with the valley at which charge is excited depending on the time separation and order of the two pulses. This effect, like the valley selection rule, has its origin in a valley phase structure, but rather than the Berry curvature it is a phase associated with the point group symmetry relations between the valleys. As such, this effect will also hold even in cases where the valley physics does not have the particular Berry curvature required for the valley selection rule, making this an effect of wide applicability that will be found in any low dimensional system possessing a valley structure to its electronic spectrum.

We first establish and explore this effect through a minimal two band tight-binding model, before employing sophisticated time dependent density functional theory (TD-DFT) calculations to show that this effect can be observed in realistic simulations of light-matter interaction in a two dimensional material, with the example of the transition metal dichalcogenide WSe$_2$. 

\section{Valley control via linear light}

A minimal model for a two dimensional semi-conductor with valley structure is provided by gapped graphene\cite{MotlaghApalkov}. This material consists of the famous honeycomb lattice of graphene, but with a sub-lattice symmetry breaking field applied such that at the K and K$^\ast$ valleys one has a gap, the size of which is controlled by the field strength.

\begin{figure*}
\includegraphics[width=0.9\textwidth]{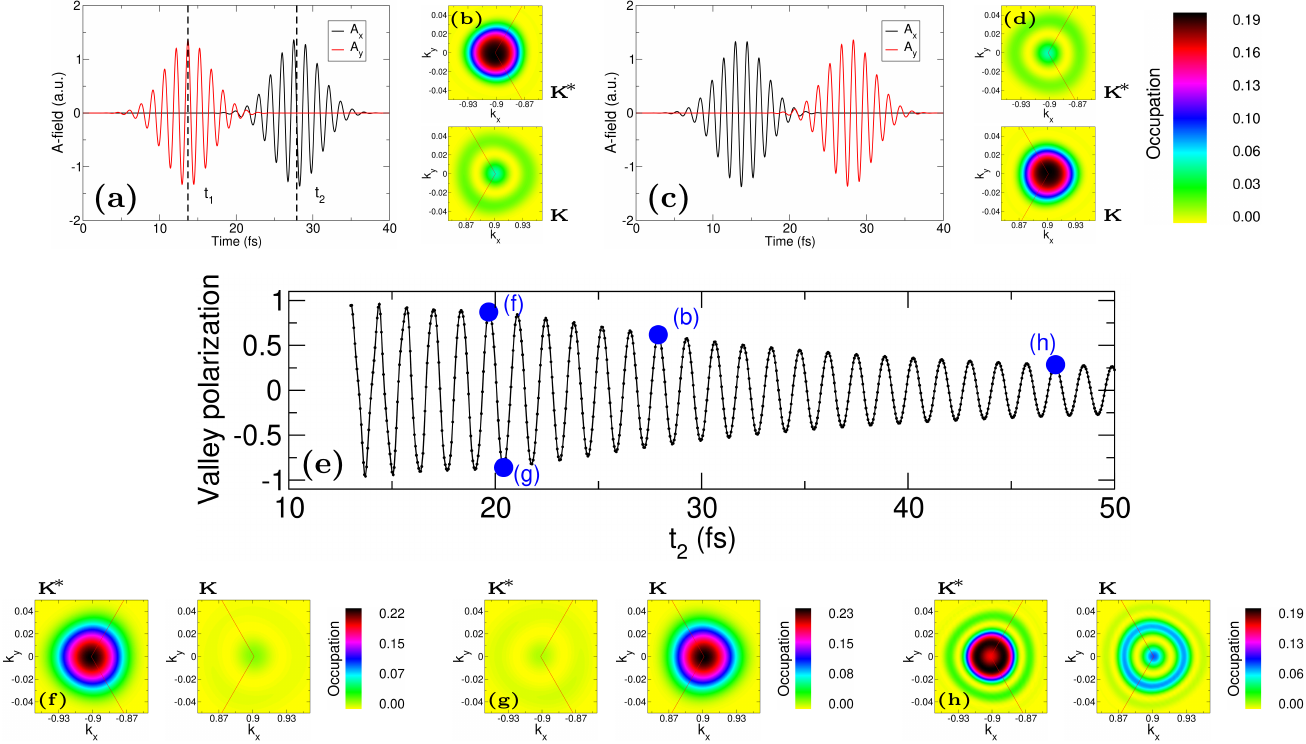}
\caption{{\it Valley selective excitation by linearly polarized light}. Two temporally separated and orthogonally polarized linear light pulses, the $y$- and $x$-polarized pulses occurring at distinct times $t_1$ and $t_2$ as shown in panel (a), result in valley distinguishing charge excitation, with charge predominately excited at the K$^\ast$ valley as shown in panel (b). Reversing the order of the pulses inverts the valley populations, with excitation now at the K valley (c,d). The valley polarisation depends in an oscillatory way on the time separation of the pulses (e), as can be seen from the valley polarization $(P_+-P_-)/(P_++P-)$ ($P_+$ and $P_-$ the excited charge after the laser pulse at the K and K$^\ast$ valleys respectively) plotted as a function of $t_2$; even for significant time separation of the pulses the effect persists. Panels (f-h) show the corresponding valley excitation in momentum space for the points indicated in panel (e).}
\label{fig:vpol}
\end{figure*}

Gapped graphene is, as in the transition metal dichalcogenides, endowed with Berry curvature of opposite sign at the K and K$^\ast$ valleys. Application of a laser pulse of circularly polarised light $\sigma^\pm$ will, therefore, excite charge at either the K or K$^\ast$ valley of the spectrum (we reproduce this standard result in Sec.~2 of the SI). Here we wish to consider the response of this system to linear light. A single linear pulse, which can be viewed as a superposition of $\sigma^+$ and $\sigma^-$ circularly polarised light, will evidently excite an equal amount of charge at all valleys, and therefore cannot "valley polarise" the material. We now show (Fig.~1) that two linear pulses result in a very different, valley discriminating, response.

To that end we consider two time separated pulses with orthogonal linear polarisation, as shown in Fig.~\ref{fig:vpol}(a). The first of these pulses is polarised in the $y$-direction with carrier envelope phase $\pi/2$ with the second in the $x$-direction with carrier envelope phase $0$. The resulting $\v k$-resolved excited charge is displayed in Fig.~\ref{fig:vpol}(b). We see that rather than couple equally to all valleys, as would be the case if a single linear pulse excited the system, a significant valley polarisation is observed with charge excited predominately at the K$^\ast$ valley. In contrast, almost no charge excited at K. If we now change the order in which we apply the two orthogonally polarized pulses of light, 
see Fig.~1(c,d), it is now the K valley is excited in place of the K$^\ast$ valley. For temporally {\it coinciding} pulse envelope maxima $t_1=t_2$ then the two linearly polarized pulses would simply represent the components of a circularly polarized pulse, with interchanging the order equivalent to changing helicity $\sigma^+\Leftrightarrow\sigma^-$ of the light which would switch valley polarisation by the valley selection rule of circular light. It is a remarkable result that when these components are temporally separated to become distinct pulses of linear light this valley polarisation switching still occurs.

To probe the dependence of valley excitation on the time separation between the two pulses, we consider a series of pulse separations $|t_2-t_1|$ and integrate the laser excited charge over crystal momentum $\v k$ to yield the charge excited at the K (+) and K$^\ast$ (-) valleys, $P_\pm$, with the "valley polarisation" then given by $(P_+-P_-)/(P_++P_-)$. This quantity encodes the valley response of the system to light and takes on the values $+1$ and $-1$ for the case of charge excited exclusively at the $K$ and $K^\ast$ valleys, while it is equal to 0 if there is no valley distinction in the response to light. In panel (e) of Fig.~\ref{fig:vpol} we show the valley polarisation as a function of the pulse envelope maxima for the second pulse, $t_2$, with the pulse envelope maxima of the first pulse, $t_1$, held fixed. A significant valley discriminating signal is seen that, moreover, oscillates as a function of the time difference between the pulses. For pulses that are not significantly time separated the valley response is close to complete valley polarization, falling to about 25\% polarization for well separated pairs of pulses. The valley distinction in the charge excitation is clearly revealed in panels (f-h) in which are displayed the $\v k$-resolved excited charge for a series of representative times $t_2$ of the second pulse (with the times indicated in the valley polarisation curve, Fig.~1(e)). As can be seen while one valley is always substantially more excited than its conjugate partner, increasing the delay between the first and second pulses both reduces the valley contrast while at the same time as resulting in somewhat more complex patterns of charge excitation

\begin{figure*}
\includegraphics[width=0.9\textwidth]{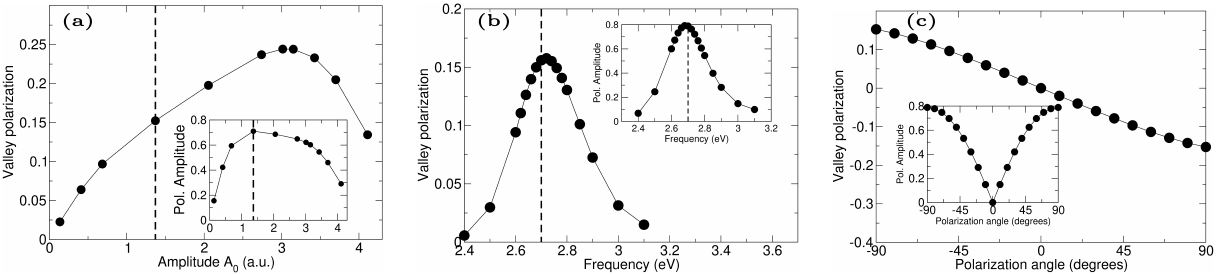}
\caption{{\it Stability of valley excitation with respect to pulse perturbations}. Stability of valley distinguishing charge excitation induced by two temporally separated and orthogonally polarized linear pulses probed as a function of changing the amplitude (a), frequency (b), and polarization angle (c) of the second pulse. In the main panel is shown the valley polarization with the pulse centre of the second pulse fixed at 40 femtoseconds, while the inset panel shows the {\it amplitude} of the valley polarization angle oscillation. As may be seen, quite substantial difference in amplitude and polarization angle can occur between the two pulses while maintaining the valley distinguishing charge excitation of the pulse pair.}
\label{fig:stab}
\end{figure*}

Thus far we have considered pulses with identical amplitude and perfectly orthogonal polarisation, and we now establish how robust this physics to non-orthogonal polarisation vectors between the pulses and non-equal vector potential amplitudes and frequencies of the pulses. Without loss of generality we can vary pulse parameters only of the second pulse. Modification of the second pulse amplitude shows that as it is reduced from that of the first (indicated by the vertical dashed line in panel Fig.~2(a)), the valley polarization is increasingly reduced, falling to zero as the second pulse amplitude vanishes. In contrast, increasing the amplitude of the second pulse results in an {\it increase} in valley contrast. This occurs as modification of the amplitude of the vector potential of the second pulse generates a reduction in amplitude of the valley polarization oscillation, but also a phase shift as a function of time difference. It is this second feature responsible for the surprising increase in valley contrast at fixed pulse time. In the inset of panel (a) is shown the amplitude of the valley oscillation curve, showing that this uniformly decreases as the two pulses become unequal in vector potential amplitude, as one would expect. Nevertheless, the valley contrast is seen to be robust to quite substantial changes in the amplitude of the second pulse. Frequency differences between the two pulses of up to 10\% (panel (c)) also generate only minor reduction of the valley polarization (the inset again shows the change in amplitude of the valley oscillation curve), with however significant changes in frequency almost completely destroying the effect. Finally in panel (c) we show the result of changing the angle between the polarization vectors of the two pulses. This strongly modifies the valley contrast, with the valley vanishing for co-linear polarization vectors (0 degrees). Again, however, for modest inaccuracies (up to 20 degrees) in orthogonality the effect is seen to be quite robust.

\begin{figure*}
\includegraphics[width=1\textwidth]{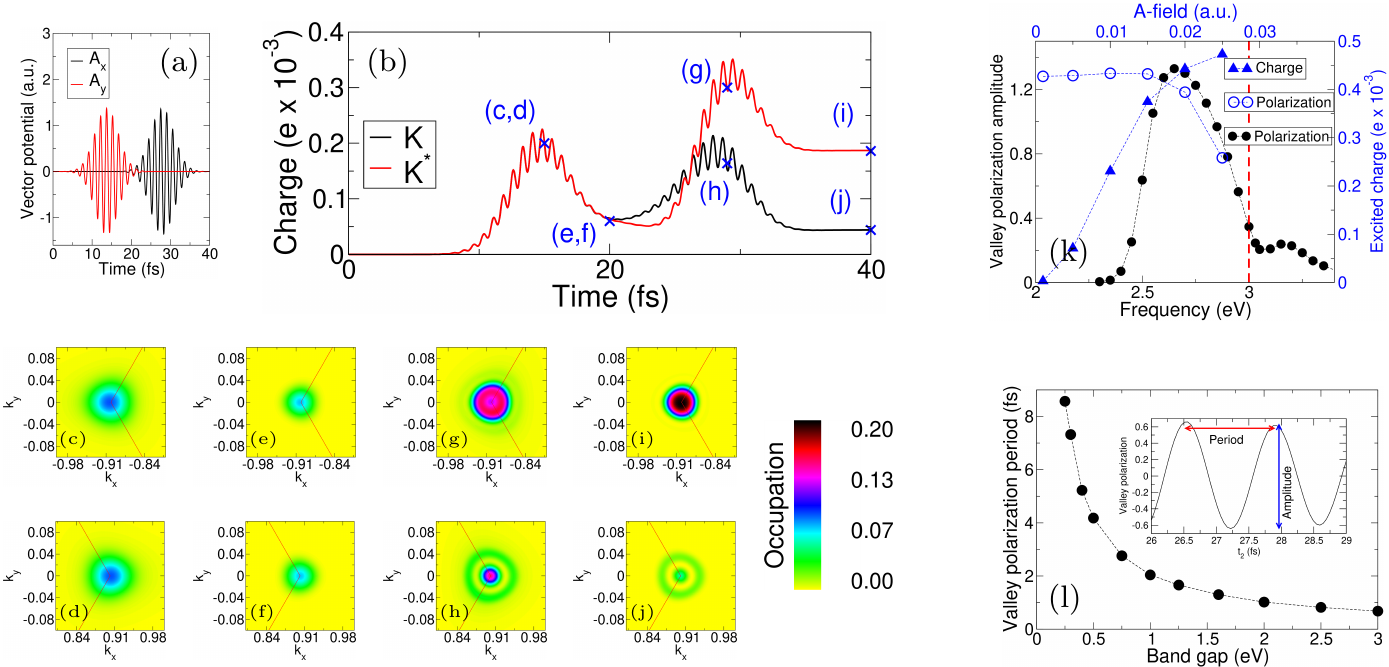}
\caption{Dynamics of charge excitation at the K and K$^\ast$ valleys for the pair of linear pulses shown in panel (a). While the first pulse excites charge equally at both valleys, see panel (b), the second linear pulse produces dramatically different dynamics with further excitation at K$^\ast$ but de-excitation at K. The momentum resolved charge excitation at each valley, shown in panels (c-j), reinforces the fact that the K valley de-excites after the second pulse. Valleys are distinguished most strongly at sub-gap frequencies, panel (k), and increasing the pulse amplitude results in an increase in excited charge but no qualitative change in the valley polarization amplitude, shown also in panel (k). The period of the valley oscillation depends sensitively on the gap, as shown in panel (l).}
\end{figure*}

To gain more insight into the origin of this effect we now consider the valley charge ($P_\pm$(t)) dynamics during laser excitation, see Fig.~3(a). Under the action of the first pulse charge is excited equally at both K and $K^\ast$, rising to a maxima at the pulse envelope peak before falling again during the pulse tail, behaviour typical of the charge dynamics induced by intense laser light in many materials. During the second pulse, however, the charge dynamics is unusual. While $P_-$ shows approximately a doubling of the excited charge at K$^\ast$, roughly what one would expect from the application of a second pulse, $P_+$ shows \emph{less} excited charge exists at the K valley: for this valley the second linearly polarized pulse de-excites. In short: while the first pulse excites at both valleys, the second pulse excites at one valley and de-excites at the conjugate valley. To highlight this excitation/de-excitation we show at various points along the $P_\pm$ curves the $\v k$ resolved excited charge, see panels (c-j). Changing the frequency and amplitude of the both pulses, see Fig.~3(k), results in changes in the strength of the effect with the amplitude of the valley polarization oscillation (defined in the inset of panel (l)) revealing a sub-gap maxima as a function of frequency, highlighting the difference with the well known selection rule for circular light which would be a maxima at the frequency of the gap. On the other hand increase in amplitude of the pulses results, as expected, in an increase of the amount excited charge but no significant change in polarization.

Finally, we consider the effect of changing the value of the band gap, for which we scale the pulse frequency such that its ratio with the gap remains unchanged. In Fig.~3(l) is shown the period of the valley polarisation oscillation as a function of gap, revealing that decrease of the gap magnitude results in a significant and non-linear increase in the oscillation period. Evidently, the evolution of the dynamical phase between the two linear pulses (which will have a rate set by the gap) plays an important role, bringing out clearly that it is the physics of wavefunction interference, i.e. coherent phase physics, destructive and constructive at the different valleys, that lies at the heart of this effect.

\section{Interference physics}

To discover the origin of this interference effect we now consider the low energy approximation to the tight-binding model of gapped graphene, the Dirac-Weyl Hamiltonian. The time dependent problem is then given by $-i \partial_t a(t) = H(\v k(t)) a(t)$ with $a(t)$ the time dependent wavefunction and

\begin{equation}
H(\v k(t)) = \begin{pmatrix}\Delta & v_F k(t) e^{i\theta(t)} \\ v_F k(t) e^{-i\theta(t)} & -\Delta\end{pmatrix}
\label{eq:DW}
\end{equation}
the Dirac-Weyl Hamiltonian (we use atomic units so $\hbar=1$). In this Hamiltonian $v_F$ is the Fermi velocity and $k(t)$ the magnitude of crystal momentum measured from the nearest K valley, the time dependence of which is given by the Bloch acceleration theorem: $\v k(t) = \v k(0) + \v A(t)/c$. The phase angle $\theta(t)$ is given by $\theta(t) = \Phi_\nu - \nu \phi(t)$ where $\nu = \pm 1$ labels the two conjugate valleys, K and K$^\ast$, and $\phi(t) = \tan^{-1} k_y(t)/k_x(t)$ is the angle of the crystal momentum measured from the valley centre. The phase $\Phi_\nu$ is additional time independent valley phase unimportant for our considerations.

In fact, for excitations within the K point valleys Eq.~\ref{eq:DW} captures almost the full dynamics of the tight-binding Hamiltonian, reflecting the accurate description that the Dirac-Weyl Hamiltonian provides near K. This Hamiltonian, however, has the great advantage of explicitly showing the phase structure in terms of the crystal momentum. To this end we note that if we are exactly at one of the K points the crystal momentum angle $\phi(t)$ is {\it time independent} for linear pulses, $\phi(t)=\phi$, being for the case of $x$- and $y$-polarisation just $\phi=0$ and $\phi=\pi/2$ respectively. This allows for the time evolution matrix to be solved exactly and, expressing the solution in terms of valence ($v$) and conduction ($c$) occupation at K, $b=(b_v,b_c)$, we can write time time propagation for a single pulse as $b_f = M(\phi) b_i$ with $b_{i,f}$ the initial and final states and $\phi$ the polarization angle of the linear light. The time propagation matrix is given by

\begin{equation}
M(\phi) = \begin{pmatrix}
C^* &
i T e^{-i\nu\phi}\\
-i T e^{i\nu\phi}&
C
\end{pmatrix}
\label{eq:prop}
\end{equation}
where the constants $C$ (complex) and $T$ (real) are given in Sec.~3 of the SI. The crucial feature of this propagation matrix is that the off-diagonal blocks, which determine charge excitation and de-excitation at K, contain the phases $\pm \nu \phi$. This shows that on charge excitation within a valley quasi-particles acquire a valley distinguishing phase combining the polarisation angle of the linear pulse ($\phi$) with a valley distinguishing sign $\nu=\pm 1$. The effect of two temporally separated pulses is given by multiplication of two of these propagation matrices, with appropriate unitary evolution of the dynamical phase during the "off" phase (duration $\delta T$) separating the pulses, $M_{total} = M_2 e^{-i\text{Diag}(\Delta_{gap},-\Delta_{gap}) \delta T} M_1$, and this immediately yields for the final conduction band occupation $|C T e^{-i\Delta \delta t}e^{-i\nu\phi_1} + C^\ast T e^{i\Delta\delta t}e^{-i\nu\phi_2}|^2$. The origin of the interference effect can be now clearly seen as in this expression we have two terms, from charge excitation during the first and second pulses, each of which has its own valley distinguishing phase arising from the different polarization vectors. Evaluating this expression reveals the interference physics driven by these two terms:

\begin{equation}
|b_c|^2 = 4|C|^2 T^2 \cos^2(\Delta \delta T - \nu (\phi_2 -\phi_1)/2)
\end{equation}
This expression captures in a microcosm the remarkable valley response found in the full dynamics using the tight-binding Hamiltonian. For orthogonal pulses $(\phi_2 -\phi_1) = \pi/2$ the K ($\nu=+1$) and K$^\ast$ ($\nu=-1$) conduction occupations are half a period out of phase and, therefore, when $K^\ast$ is excited $K$ is de-excited and vice versa, the basic result shown Fig.~1. The peculiar valley switching on changing the order of the pulses is now also easily explained: switching the order of the $x$- and $y$-polarized pulses sends $(\phi_2 -\phi_1)$ from $\pi/2$ to $-\pi/2$, thus switching the the valley at which charge is excited. Similarly, the cosine-like dependence on the angle of the second pulse, see Fig.~2(c) inset, is also reproduced by this result. 

The valley discriminating response to pairs of orthogonally polarised linear light thus arises from interference between the excitations of each pulse, as quasiparticles, in the vicinity of the valleys have phases controlled by both the pulse polarisation vector and the valley index. While we have shown this exactly at the high symmetry K points, it will evidently hold in the region of these valleys and indeed this is seen to be so in numerical examination (see SI Sec.~3).

Thus far we have considered only single particle excitations, and we now turn to the question of whether composite many-body excitations (such as excitons and trions\cite{hanbicki_high_2016}) will inherit this phase structure. If this is to be the case, all single particle excitations must possess the phase structure identified above as resulting from a complete laser pulse cycle. In fact, it is easy to show that at any time {\it during} the pulse all possible transitions between valence and conduction will have exactly the valley distinguishing phase identified above, see SI. If we then write an exciton wavefunction as

\begin{equation}
\ket{\phi_X} = \sum_{vc\v k} A^X_{vc} c_{c\v k}^\dagger c_{v\v k} \ket{0}
\end{equation}
we see that for excitations by linear polarised light all amplitudes $A^X_{vc}$ will have the valley discriminating phase $\nu\phi$ and, therefore, the exciton wavefunction itself will bear an imprint of the polarization vector via this phase. Whether intereference between the excitonic fraction occurs is a more subtle question, requiring solution of time dependent many-body problem, which we do not consider here.

\section{The effect in WSe$_2$}

Any interference effect requires for its manifestation coherent charge dynamics which typically holds for the ultrafast femtosecond scale pulses we consider here. However any system in which this effect may occur, such as transition metal dichalcogenides, will possess time dependent wavefunctions with hugely more degrees of freedom than the simple tight-binding model considered here, as well as a considerably more complex band structure. All of this can alter the coherent dynamics of the wavefunction and so modify (or destroy) the interference physics that drives valley selection. This possibility must be carefully explored in order to establish the effect we propose as realistic, and we thus now perform time dependent density function theory simulations for the transition metal dichalcogenide WSe$_2$. For all calculations employ the adiabatic LDA functional, an approach that has been shown to be highly accurate in treating very early time spin and charge dynamics in many materials. For complete numerical details we refer the reader to the SI, Sec.~5.

\begin{figure*}
\includegraphics[width=0.9\textwidth]{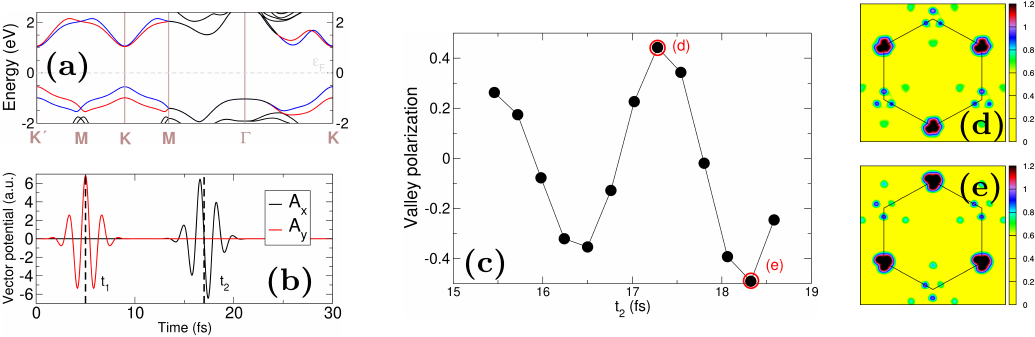}
\caption{Valley interference effect in WSe$_2$. Shown in panel (a) is the low energy band structure with blue (red) indicating spin up (down) bands. Employing single cycle pairs of orthogonally polarised linear pulses (see panel (b)) we find exactly the same phenomena of valley control by pulse separation, see panels (c-e), as found in the gapped graphene model.}
\end{figure*}

The low energy band structure is shown in Fig.~4(a), in which the characteristic strong spin orbit spin split bands can be seen near K and K$^\ast$. We consider two single cycle pulses of FWHM 2.8~fs and orthogonal polarisation (see Fig.~4(b)) and, as for the model calculations, change the envelope peak of the second pulse, $t_2$, while holding that of the first pulse fixed. In panel (c) we show the valley polarization, now integrated in the valleys defined at K and K$^\ast$, and, just as in the model calculation, an oscillation of the valley polarization is again observed. Inspection of the $\v k$-resolved momentum space excitation reveals the valley discriminating charge excitation seen in the gapped graphene calculations. Note that here the excitation is spin polarized due to the spin-split valence bands. This result is perhaps not as surprising as it may appear, as valley phase structures are generally the most robust aspect of the simple models used to describe 2d materials such as graphene and the dichalcogenides.

\section{Conclusions}

We have shown that the paradigm that circular polarised light offers the only route to control the valley degree of freedom does not hold at femtosecond time scales in which ultrafast spin and charge dynamics are coherent. Pairs of orthogonally polarised yet temporally separated linear pulses selectively excite conjugate valleys depending on the duration between pulses and the order of the pulses (i.e., $x$- followed by $y$- polarisation or vice versa). While the circularly polarised light selection rule relies on Berry curvature of an avoided band crossing, this effect depends on coherent interference between the unique valley phases that an avoided crossing is endowed with. These valley distinct phases are governed by the point group relationships between the various electronic valleys and thus this effect should be general to all 2d materials with the valley degree of freedom, a wide class of materials. In materials with spin spit valence bands this effect will result in spin-valley locking, just as in the case of circular light.

Evidently the requirement for coherent dynamics renders the effect vulnerable to the phase breaking that can occur outside the femtosecond regime, either from lattice dynamics (typically $> 50$~fs) or coulomb interaction scattering\cite{pashalou_coherent_2020} (typically $> 10-200$~fs). While we have established and understood this new route to valley control on the basis of a simple gapped graphene model, we have confirmed the effect in sophisticated TD-DFT calculations that have, by now, been established as the gold standard for very early time spin and charge dynamics. The pulse parameters are not onerous, with the 7~fs FWHM easily achievable, however careful femtosecond scale control over the pulse maxima is required. 

Our model calculations do not include the Coulomb interaction and our TD-DFT calculations, while capturing many effects of interaction, do not include a treatment of the dynamics of excitons, which remain challenging for the \emph{ab-initio} approach. However, on general grounds we have shown that composite many body particles (such as excitons) will inherit the same valley phase structure found for single particle valence to conduction transitions, thus allowing for valley discriminating interference effects beyond the single particle regime. The effect we propose has, moreover, been established on general grounds and probed over a wide range of pulse parameters. We thus expect that this effect could be observed in experiment, opening new routes towards spin-valley-orbitronics with linearly polarized light at the femtoscale.


\newpage
\section{Supplementary information}

\subsection{Tight-binding simulation}
\label{SI:TB}

\begin{figure*}[b!]
\includegraphics[width=0.90\textwidth]{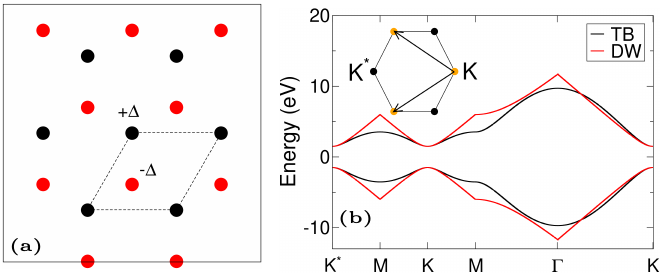}
\caption{Panel (a): Lattice of graphene showing the two interpenetrating hexagonal lattice of the honeycomb lattice (dark and ligh shaded circles) along with the gap opening field on each sublattice $\pm\Delta$. (b) Band structure of gapped graphene using the tight-binding Hamiltonian (Eq.~\ref{eq:TB}) and the low-energy Dirac-Weyl approximation (Eq.~\ref{eq:siDW})
}
\label{fig:S0}
\end{figure*}

The general two-centre tight-binding Hamiltonian is given by

\begin{equation}
H_0 = \sum_{ij} t_{ij} c^\dagger_j c_i + \sum_i \Delta_i c_i^\dagger c_i 
\end{equation}
where $t_{ij}$ is the hopping amplitude between sites $i$ and $j$ (we suppress all other atomic indices). The field $\Delta_i$ represents an on-site field that, in the case of gapped graphene, alternates sign between the two sublattices. Time evolution is through the Sch\"odinger equation

\begin{equation}
i \partial_t \ket{\Psi(t)} = H \ket{\Psi(t)}
\end{equation}
where $\ket{\Psi(t)}$ is the system ket at time $t$ for crystal momentum $\v q$ at $t=0$. In this expression we employ the length gauge and so we have

\begin{equation}
H = H_0 + \v r.\v E(t)
\label{eq:S2}
\end{equation}
Time dependent system ket can be expanded in a basis of Bloch states at crystal momentum $\v k(t)$

\begin{equation}
\ket{\Psi_{\v q}(t)} = \sum_{\alpha} c_{\alpha \v q}(t) \ket{\Phi_{\alpha \v k(t)}}
\label{eq:S1}
\end{equation}
with $\v k(t) = \v q + \v A(t)/c$ given by the Bloch acceleration theorem. The Bloch states $\ket{\Phi_{\v k \alpha}}$ are given by

\begin{equation}
\ket{\Phi_{\v k \alpha}} =  \frac{1}{\sqrt{N_n}} \sum_{\v R_i} e^{i\v k.(\v R_i^{(n)} + \m \nu_\alpha^{(n)})}
\ket{\v R_i + \m \nu_\alpha}
\end{equation}
where $\ket{\v R_i + \m \nu_\alpha}$ is a localised orbital at site $\v R_i + \m \nu_\alpha$ with $\v R_i$ a lattice vector and $\m \nu_\alpha$ the basis vector of sub-lattice $\alpha$.

Taking the time derivative of Eq.~\ref{eq:S1} we then find

\begin{equation}
i \partial_t \ket{\Psi_{\v q}(t)} = i
\sum_\alpha \left( \dot{c}_{\alpha\v q}(t) \ket{\Phi_{\alpha \v k(t)}} 
+c_{\alpha\v q}(t) \dot{\v k}.\m\nabla_{\v k} \ket{\Phi_{\alpha \v k(t)}} \right) 
\end{equation}
and employing Eq.~\ref{eq:S2} leads to

\begin{equation}
i\hbar \dot{c}_{\alpha'\v q} = 
\sum_\alpha \mel{\Phi_{\alpha' \v k(t)}}{H_0}{\Phi_{\alpha \v k(t)}} c_{\alpha\v q}
+\v E(t).\mel{\Phi_{\alpha' \v k(t)}}{\left(\v r - i\m\nabla_{\v k}\right)}{\Phi_{\alpha \v k(t)}} c_{\alpha\v q}
\end{equation}
which simplifies to

\begin{equation}
i \partial_t c_{\v q} = H_0(\v k(t)) c_{\v q}
\label{eq:cSE}
\end{equation}
the equation of motion for $c$ is the 2-vector of expansion coefficients in the basis states at $\v k(t)$.


{\bf Tight-binding description of gapped graphene}: Gapped graphene consists of the honeycomb lattice of graphene, with a "mass generating" field $+\Delta_{gap}$ and $-\Delta_{gap}$ on the two sub-lattices, see Fig.~\ref{fig:S0}. Employing the simplest possible model we have in a sub-lattice Bloch basis the well known Hamiltonian

\begin{equation}
H_{TB} = \begin{pmatrix}
\Delta_{gap} & t_{hop} f(\v k) \\ t_{hop} f^\ast(\v k) & -\Delta_{gap} \end{pmatrix}
\label{eq:TB}
\end{equation}
where $f(\v k) = \sum_j e^{i\v k.\m \nu_j}$ is the Bloch lattice sum over nearest neighbours $\m \nu_j$. This Hamiltonian depends on only two parameters: the nearest neighbour tight-binding parameter $t_{hop}$ describing electron hopping between the two inter-penetrating sublattices of the honeycomb lattice, and $\Delta_{gap}$ that controls the magnitude of the gap at the K point ($E_{gap} = 2\Delta_{gap}$).


{\bf Pulse waveform}: In all calculations the laser pulse is described by Gaussian envelope form:

\begin{equation}
\v A(t) = \v A_0 \exp\left(-\frac{(t-t_0)^2}{2\sigma^2}\right) \cos(\omega t + \phi)
\end{equation}
where $\v A_0$ is the pulse amplitude vector, $\sigma$ is related to the full width half maximum by $FWHM = 2\sqrt{2\ln{2}}\sigma$, $\omega$ is the frequency of the light and $\phi$ the carrier envelope phase.

\begin{figure*}
\includegraphics[width=0.90\textwidth]{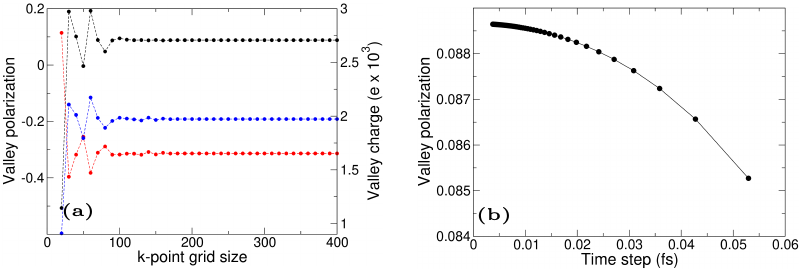}
\caption{Panel (a): Convergence with respect to k-mesh of valley charge and polarisation after laser pulse excitation. (b) Convergence with respect to the time step. In both cases the laser pulse consists of two linearly polarized and time separated waveforms as shown in panel Fig.~1(b) of the manuscript.
}
\label{fig:S1}
\end{figure*}

{\bf Numerical method}: We solve Eq.~\ref{eq:cSE} for the case of the gapped graphene Hamiltonian via the Crank-Nicolson method:

\begin{equation}
c_{n} = \left(1-\frac{H_0(t_{n})}{2i}\right)^{-1} 
\left(1+\frac{H_0(t_{n-1})}{2i}\right) c_{n-1}
\end{equation}

A k-grid of $400\time400$ points in the Brillouin zone is generally employed (convergence of the valley polarisation and change is shown in Fig.~\ref{fig:S1}a). The necessity of this high k-point convergence arises as the charge excitation is localised at the high symmetry K points and often exhibits fine structures; this turns out to be particular important for the convergence of the current density. For the time step of the Crank-Nicholson equation we find a rather coarse graining is often sufficient, reflecting the limited number of degrees of freedom in the minimal tight-binding model that we employ. The convergence with respect to time step is shown in Fig.~\ref{fig:S1}b.

\subsection{Valley selection rule in gapped graphene}

\begin{figure*}
\includegraphics[width=0.95\textwidth]{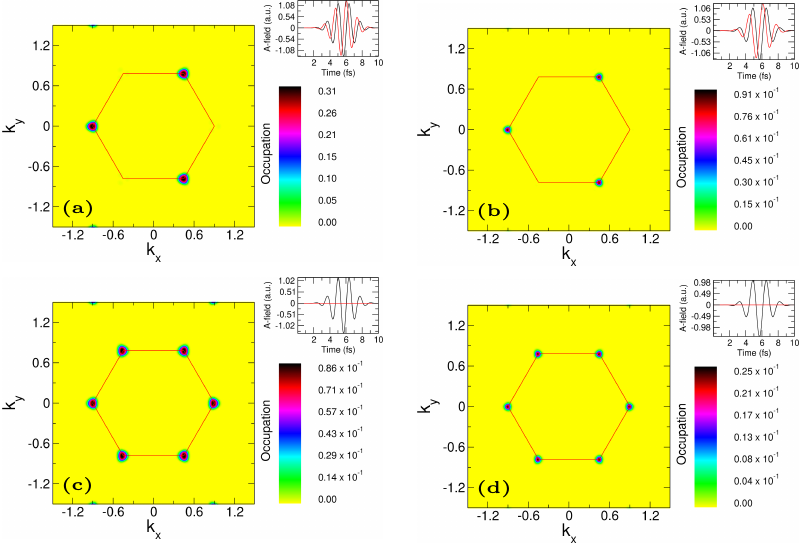}
\caption{{\it Valley selection rule of gapped graphene}. The topmost panels (a) and (b) display the response to circularly polarized light $\sigma^-$ which excites at the K$^\ast$ valley exclusively. Shown are light pulses at the gap frequency (3~eV) and at a sub-gap frequency of 2.7~eV. Correspondingly linear pulses are shown in panels (c) and (d), for the same frequencies, with charge now excited at all valleys equally. Black and red lines denote the $A_x$ and $A_y$ components of the vector potential.
}
\label{fig:val}
\end{figure*}

The selection rule for circularly polarized light arises as the K and K$^\ast$ valleys are endowed with Berry curvature of opposite sign. For light frequency equal to the optical gap significant charge is excited, panel Fig.~\ref{fig:val}(a) while for sub-gap frequencies the excited charge is significantly reduced Fig.~\ref{fig:val}(b). Linearly polarized pulses, which can decomposed into two equal amplitude $\sigma^+$ and $\sigma^-$ pulses, will not distinguish valleys in the optical response, Fig.~\ref{fig:val}(c,d) for the same gap and sub-gap frequency pulses respectively.

\subsection{Derivation of valley selective interference}

\begin{figure*}
\includegraphics[width=0.95\textwidth]{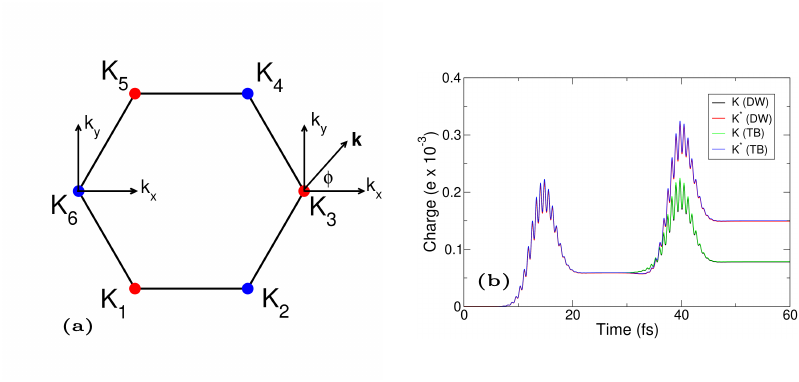}
\caption{Panel (a): Brillouin zone indicating the two stars of high symmetry points K and K$^\ast$, K$_1$ and K$_2$ respectively. Shown also are the local coordinate systems that the Dirac-Weyl Hamiltonian is referred to; the angle $\phi$ is the azimuthal angle of crystal momentum that plays a key role in the dynamics. Panel (b): Time evolution of charge occupation at the K and K$^\ast$ valleys for dynamics governed by the tight-binding Hamiltonian (Eq.~\ref{SI:TB}) and Dirac-Weyl low energy Hamiltonian (Eq.~\ref{eq:siDW}).
}
\label{fig:model}
\end{figure*}

Our derivation will be based on the gapped Dirac-Weyl equation that describes electronic structure in the vicinity of the K points of gapped graphene:

\begin{equation}
H_{DW} = \begin{pmatrix}
\Delta_{gap} & v_F k e^{i\theta} \\  v_F k e^{-i\theta} & -\Delta_{gap}
\end{pmatrix}
\label{eq:siDW}
\end{equation}
In this expression $\Delta$ determines the band gap at the K points ($E_{gap} = 2\Delta_{gap}$), $v_F$ the Fermi velocity, $k$ the magnitude of the crystal momentum measured from the K point for which Eq.~\ref{eq:siDW} describes the electronic structure, and $\theta = \Psi_\nu - \nu \phi$, with $\nu=+1$ at the K points and $\nu=-1$ and the K$^\ast$ points and $\phi = \tan^{-1} k_y(t)/k_x(t)$ the azimuthal angle of the crystal momentum $\v k$ (illustrated in Fig.~\ref{fig:model}(a)). The phase $\Psi_\nu$ has no time dependence as it simply encodes the $C_3$ symmetry that the two sets of K points possess (it takes on values $-2\pi/3$, $2\pi$, and $2\pi/3$ for K points 1, 3, and 5 and values $\pi/3$, $-\pi/3$, and $\pi$ for K points 2, 4, and 6 (see Fig.~\ref{fig:model}(a)). This phase plays no role in the valley selective interference effect we will now discuss.

For the dynamics near the K valleys the Dirac-Weyl equation reproduces very closely the results of the tight-binding model, see Fig.~\ref{fig:model}(b).

The equation of motion for time evolution in the pseudospin basis in which Eq.~\ref{eq:siDW} is expressed is formally the same at that of the time evolution of the tight-binding Hamiltonian in the atomic sub-lattice basis derived in Sec.~\ref{SI:TB}:

\begin{equation}
i\partial_t c(t) = H_{DW}(\v k(t)) c(t)
\label{eq:SWDW1}
\end{equation}
where as before the time evolution of the crystal momentum $\v k$ is given by the Bloch acceleration theorem,

\begin{equation}
\v k(t) = \v k(0) + \v A(t)/c
\end{equation}
with $\v k(0)$ the crystal momentum before the pulse, and $c(t)$ is the time dependent wavefunction in the pseudospin basis.

We now consider a general pulse and break the time evolution into discrete segments

\begin{equation}
\v k_i = \v k_0 + \frac{\v A_i}{c}
\end{equation}
where the segment index $i$ runs $i=0,N$.

The time evolution matrix in the atomic basis is given by

\begin{equation}
Q = \text{exp}\left(-i\int_0^t dt'\, H(t')\right)
\end{equation}
with $T$ the time ordering operator. This matrix evidently relates the wavefunction (in the atomic basis) from the initial time $t=0$ to the final time $t$.

\begin{equation}
b_f = Q b_i
\end{equation}
$Q$ will be composed of a series of matrices describing the different segments:

\begin{equation}
Q = Q_N\ldots Q_0
\end{equation}
where in each segment the vector potential is given by $\v A_i$ and the corresponding Hamiltonian for the segment time evolution is $H_{DW}(\v k_i)$.

We now transform to a Houston basis in which the time dependent wavefunction is expanded in the two eigenstates of the Dirac-Weyl equation at $\v k(t)$. The unitary transformation to achieve this is just the column matrix of the eigenvectors of Eq.~\ref{eq:DW} given by

\begin{equation}
U(k,\theta) = \frac{1}{\sqrt{2}}\begin{pmatrix}
f_- & f_+ \\ -f_+ e^{-i\theta} & f_- e^{-i\theta}
\end{pmatrix}
\end{equation}
where $f_\pm = \sqrt{1\pm \Delta_{gap}/\sqrt{(v_F k)^2 + \Delta_{gap}^2}}$. This transforms the time dependent pseudospin wavefunction $b(t)$ to a time dependent wavefunction expressed in the Houston basis $c(t)$:

\begin{equation}
c(t) = U^\dagger(k,\theta) b(t)
\end{equation}

Applying this transformation to the time propagation matrix $Q$ then yields

\begin{equation}
M = U_N^\dagger P_N\ldots P_0 U_0
\label{eq:Qseq}
\end{equation}
where the matrix $P_i$ is given by

\begin{equation}
P_i = 
\begin{pmatrix}
C_i &
i T_i e^{i\theta}\\
-i T_i e^{-i\theta}&
C_i^\ast
\end{pmatrix}
\label{eq:P}
\end{equation}
where

\begin{equation}
C = \frac{1}{2}\left(f_+^2 e^{-iE_i \Delta t} + f_-^2 e^{iE_i \Delta t}\right)
\end{equation}
and

\begin{equation}
T = f_+ f_- \sin(E_i \delta t)
\end{equation}
Note that $E_i$ is the energy of the corresponding time evolution segment at constant $\v k_i$, $E_i = \sqrt{(v_F k_i)^2 + \Delta_{gap}^2}$ (similarly, the $f_\pm$ are evaluated at $\v k_i$).

An important feature of Eq.~\ref{eq:P} is that it satisfies form preservation under composition, i.e., $P_1 P_2 = P_{12}$ with $P_{12}$ another matrix identical in form to Eq.~\ref{eq:P} but with altered coefficients. Explicitly:

\begin{equation}
\begin{pmatrix}
C_1 &
i T_1 e^{i\theta}\\
-i T_1 e^{-i\theta}&
C_1^\ast
\end{pmatrix}
\begin{pmatrix}
C_2 &
i T_2 e^{i\theta}\\
-i T_2 e^{-i\theta}&
C_2^\ast
\end{pmatrix} =
\begin{pmatrix}
(C_1 C_2 - T_1T_2) &
i (C_1T_2 + T_1 C_2^\ast) e^{i\theta}\\
-i (C_1T_2 + T_1 C_2^\ast)^\ast T e^{-i\theta}&
\left(C_1 C_2 - T_1T_2\right)^\ast
\end{pmatrix}
\end{equation}
This crucial feature of the time evolution then allows Eq.~\ref{eq:Qseq} for time full time evolution matrix to be straightforwardly evaluated. At the end points of the time evolution we have $\v k_0 = \v k_N = \v 0$ (measured from the high symmetry expansion points K or K$^\ast$), and so:

\begin{equation}
U_N^\dagger = \begin{pmatrix}
0 & -e^{i\theta_0} \\ 1 & 0
\end{pmatrix}
\text{, and}\,\,
U_0^\dagger = \begin{pmatrix}
0 & 1 \\ -e^{-i\theta_0} & 0
\end{pmatrix}
\end{equation}

We then find for the time evolution of a general single cycle pulse in the Houston basis

\begin{equation}
M(\phi) = 
\begin{pmatrix}
C^\ast & i T e^{i\nu(\phi-\phi_0)} \\
-i T e^{-\nu(\phi-\phi_0)} & C
\end{pmatrix}
\end{equation}
where we have used $\theta = \Psi_\nu - \nu \phi$ and note that $\phi_0$ is the azimuthal angle of the initial point and $\phi$ the polarization angle of the linear pulse.

As discussed in the main text, the crucial feature of this propagation matrix is the presence in the off-diagonal blocks of the phase $e^{i\nu(\phi-\phi_0)}$ that combines (i) valley discriminating physics through $\nu$, (ii) the polarisation angle of the linear light through $\phi$ and, less importantly for our considerations, (iii) the azimuthal angle of the initial state before the laser is applied $\phi_0$. (For dynamics at the K point this is undefined except in the limit of fixed $\phi_0$ and then $k\to 0$.)

The dynamics of two pulses evidently requires multiplication of two of these single pulse propagation matrices with the appropriate unitary evolution of the dynamical phase between the pulses. This can be written as

\begin{equation}
M_{12} = M(\phi_2)
\begin{pmatrix}
e^{-i\Delta_{gap} \delta T} & 0 \\
0 & e^{i\Delta_{gap} \delta T}
\end{pmatrix}
M(\phi_1)
\label{eq:Mform}
\end{equation}
and the situation we are interested in is the case of time evolution from the ground state at K to a final excited state

\begin{equation}
\begin{pmatrix} c_v \\ c_c \end{pmatrix} =
M_{12} \begin{pmatrix} 1 \\ 0 \end{pmatrix}
\end{equation}
Evaluation of Eq.~\ref{eq:Mform} yields

\begin{equation}
M_{12} = \begin{pmatrix}
A & i B \\ -i B^\ast & A^\ast
\end{pmatrix}
\end{equation}
where

\begin{eqnarray}
A & = & (C_1 C_2)^\ast e^{-\Delta_{gap} \delta T} + T_1 T_2  e^{i\nu(\phi_2-\phi_1)} e^{\Delta_{gap} \delta T} \\
B & = & C_2^\ast T_1 e^{i\nu(\phi_1-\phi_0)} e^{-\Delta_{gap} \delta T} + C_1 T_2  e^{i\nu(\phi_2-\phi_0)} e^{\Delta_{gap} \delta T}
\end{eqnarray}

This expression encodes the interference of excitation by two pulses with each transition endowing the excited quasiparticles with a phase ($e^{i\nu\phi_{1,2}}$) combining the valley degree of freedom and the polarization angle of the linear light.

For pulses identical save for the polarization vector, these can interfere destructively or constructively which can be seen clearly by considering the probability for excitation $|c_c|^2$

\begin{equation}
|c_c|^2 = 4(1-|T|)^2 T^2 \cos^2(\Delta \delta T - \nu (\phi_2 -\phi_1)/2).
\end{equation}
This is the expression derived in the paper and reveals that the excitation if half a period out of phase at the two valleys $\nu=\pm1$; when one valley is excited (constructive interference) the other valley will be de-excited (destructive interference).

\subsubsection{Excitation during the pulse and away from the K points}

\begin{figure*}[t!]
\includegraphics[width=0.95\textwidth]{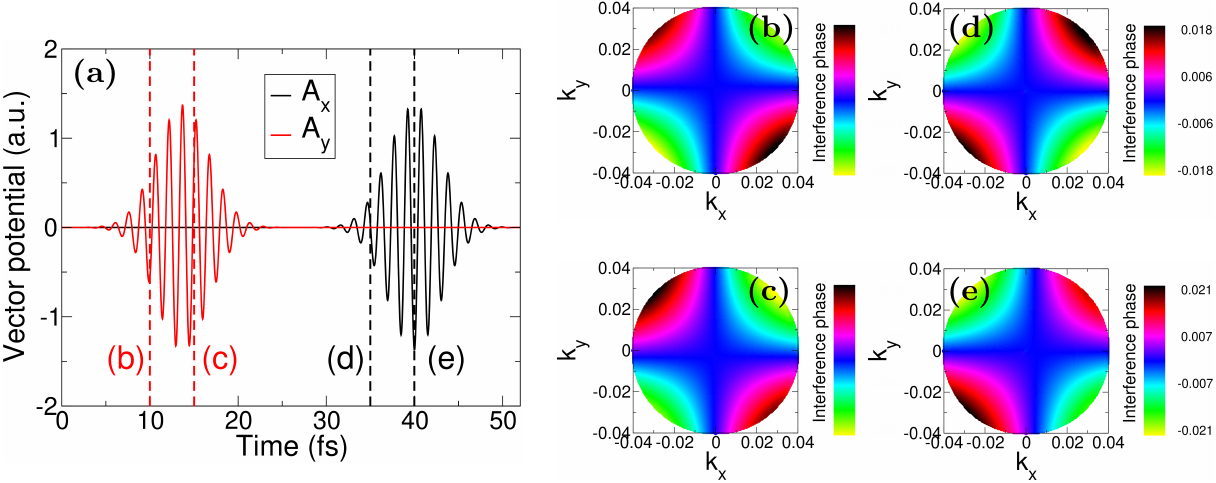}
\caption{Valley interference phase in the vicinity of the K point. Panel (a): pulse waveform showing two temporally separated linearly polarized pulses. Panels (b-c): the valley distinguishing interference phase showing that in the vicinity of K and K$^\ast$ this phase is nearly equal to $\pi$ for the first pulse (b,c) and nearly equal to $0$ for the second pulse. This difference of $\pi$ between the two valleys results in destructive interference at one valley and constructive interference at the conjugate valley.
}
\label{fig:phase}
\end{figure*}

To show that the phase feature of the full (single cycle) pulse also holds for any excitation {\it within} the pulse we must replace the end-point unitary matrix by the more general form that holds at finite $\v k$. Evaluation of this then yields

\begin{eqnarray}
M(t,\phi) & = & 
\begin{pmatrix}
f_- & f_+ \\ -f_+ e^{-i\theta} & f_- e^{-i\theta}
\end{pmatrix}
\begin{pmatrix}
C &
i T e^{i\theta}\\
-i T e^{-i\theta}&
C^\ast
\end{pmatrix}
\begin{pmatrix}
0 & 1 \\ -e^{-i\theta_0} & 0
\end{pmatrix} \\
& = &
\begin{pmatrix}
(f_+ C^\ast - i f_- T) e^{i(\theta-\theta_0)} & -i f_+ T + f_- C \\
-(i f_+ T + f_- C^\ast) e^{i(\theta-\theta_0)} & f_+ C + i f_- T
\end{pmatrix}
\end{eqnarray}
showing that the time evolution matrix for excitations during the pulse show exactly the same phase structure at each time $t$ that are found for the time evolution matrix of the complete pulse.

Finally we address the question of how this result {\it at} the K valleys holds {\it in the vicinity} of the K valleys. In the derivation above a key role is played by the fact that exactly at the K points then the phase of the Hamiltonian is exactly the polarisation angle of the linear pulse. Evidently, this is expected to hold approximately provided one does not displace the $\v k$-vector of interest too far from one of the high symmetry K points. That this is so can be checked by, from the Crank-Nicolson method, numerically obtaining the time evolution matrix $M(\phi)$ for each of the two linear pulses at a $\v k$-vector distant from one of the high symmetry K points. From these two evolution matrices can then be determined the phase difference of the off-diagonal blocks for the same $\v k$ vector displacement from K and K$^\ast$. For $\v k = \v 0$ this is exactly equal to $\pi$ for the first pulse and exactly equal to $0$ for the second pulse, as shown above, and, as can be seen in Fig.~\ref{fig:phase} near K it is, as expected, very close to $\pi$ for the first pulse and very nearly equal to $0$ for the second pulse.

\subsection{Rectangular pulse model of valley selective interference}

\begin{figure*}
\includegraphics[width=1\textwidth]{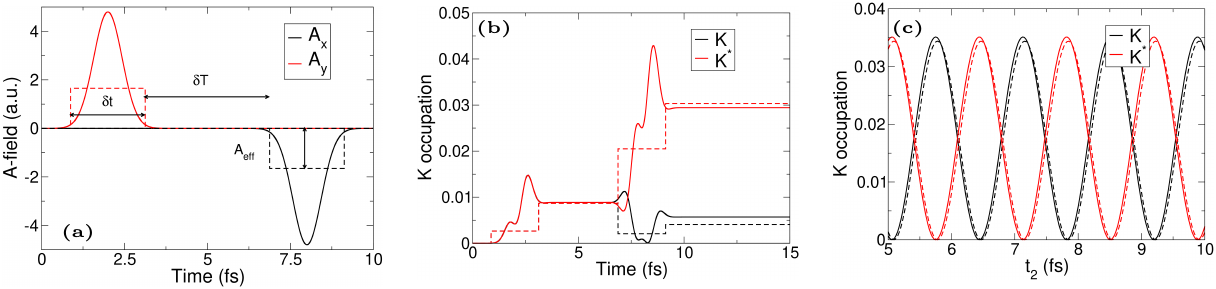}
\caption{{\it Charge dynamics of excitation and de-excitation at the high symmetry K points}. Shown in panel (a) are two linearly polarized pulses of Gaussian envelope and two pulses with rectangular envelope, with the latter pulse parameters (amplitude $A_{eff}$ and duration $\delta t$) chosen so that the charge dynamics most closely matches those of the Gaussian envelope pulse pair. The close agreement can be seen in panels (b) in which is shown the K point occupation as a function of time for these two sets of pulses. Changing the pulse centre of the second pulse results is a continuous variation in excited charge at K and K$^\ast$ which is very well captured by the rectangular pulse.}
\label{fig:cd}
\end{figure*}

Our derivation will again be based on the gapped Dirac-Weyl equation that describes electronic structure in the vicinity of the K points of gapped graphene, Eq.~\ref{eq:DW}. The time evolution equation expressed in the pseudospin basis is formally the same at that of the time evolution of the tight-binding Hamiltonian in the atomic sub-lattice basis derived in Sec.~S1:

\begin{equation}
i\partial_t c(t) = H_{DW}(\v k(t)) c(t)
\label{eq:SWDW}
\end{equation}
where as before the time evolution of the crystal momentum $\v k$ is given by the Bloch acceleration theorem,

\begin{equation}
\v k(t) = \v k(0) + \v A(t)/c
\end{equation}
with $\v k(0)$ the crystal momentum before the pulse, and $c(t)$ is the time dependent wavefunction in the pseudospin basis.

We now consider a rectangular pulse as shown in Fig.~\ref{fig:cd}(a) and the dynamics at one of the K points, i.e. $\v k(0) = \v K$. The dynamics of charge excitation at the K and  K$^\ast$ points for these two pulses (i.e. rectangular and Guassian envelopes) are in fact remarkably similar, see Fig.~\ref{fig:cd}(b). The dynamics of the rectangular pulse evidently proceeds via a series of jumps, and does not capture transient excitation associated with the peak of the Gaussian envelope, but the final K point conduction occupations after both the first and second pulse are very similar to those that result from the Gaussian pulse. Going further one can vary the time of the second pulse, $t_2$, and, as is shown in Fig.~\ref{fig:cd}(c) the oscillation of the K and K$^\ast$ conduction occupations is almost identical for the two pulse forms. As the rectangular pulse captures the K point charge dynamics, and most importantly the valley distinguishing oscillation with pulse separation, we can therefore use this simplified pulse to understand in a simplified way the origin of this effect

Evidently, the azimuthal angle of the crystal momentum, $\phi(t) = \tan^{-1} k_y(t)/k_x(t)$, remains constant during the dynamics, which evidently is true for the dynamics of any linear pulse at the K point. The use of a rectangular pulse then ensures that, in addition to this, the time dependence of the magnitude of the crystal momentum $k(t)$ evolves through a series of discrete jumps, being equal to $A_{eff} = |\v A_{eff}|$ during a rectangular pulse ($\v A_{eff}$ is the amplitude of the rectangular pulse, see Fig.~\ref{fig:cd}(a)) and zero otherwise.

This evolution by discrete jumps implies that Eq.~\ref{eq:SWDW} can be integrated directly to give a time propagation matrix

\begin{equation}
Q = \text{exp}\left(-i\int_0^t dt'\, H(t')\right)
\end{equation}
with $Q$ relating the pseudospin wavefunctions before and after the pulse:

\begin{equation}
b_f = Q b_i
\end{equation}
$Q$ will be composed of a series of matrices describing the different time intervals during which the vector potential is either 0 or $\v A_{eff}$.
It is now useful to transform to a Houston basis in which the basis for the time dependent wavefunction are the two eigenstates at $\v k(t)$. The unitary transformation to achieve this is just the column matrix of the eigenvectors of Eq.~\ref{eq:DW} given by

\begin{equation}
U(k,\theta) = \frac{1}{\sqrt{2}}\begin{pmatrix}
f_- & f_+ \\ -f_+ e^{-i\theta} & f_- e^{-i\theta}
\end{pmatrix}
\end{equation}
where $f_\pm = \sqrt{1\pm \Delta_{gap}/\sqrt{(v_F k)^2 + \Delta_{gap}^2}}$. This transforms the time dependent pseudospin wavefunction $b(t)$ to a time dependent wavefunction expressed in the Houston basis $c(t)$:

\begin{equation}
c(t) = U^\dagger(k,\theta) b(t)
\end{equation}

Applying this transformation to the time propagation matrix $Q$ yields

\begin{equation}
M(\phi) = U^\dagger(K,\theta_0) U(k_{eff},\theta_i) 
\begin{pmatrix}
e^{-i E_{eff} \delta t} & 0 \\
0 & e^{i E_{eff} \delta t}
\end{pmatrix}
U^\dagger(k_{eff},\theta_i) U(K,\theta_0)
\end{equation}
where $E_{eff} = \sqrt{(v_F A_{eff})^2 + \Delta_{gap}^2}$ with $A_{eff}$ the amplitude of the rectangular envelope (see Fig.~S2(a)), and we have renamed the propagation matrix $M(\phi)$ and explicitly indicated its dependence on the polarisation vector of the linear light through the angle $\phi$. This expression can be evaluated to give

\begin{equation}
M(\phi) = \begin{pmatrix}
C &
-i T e^{-i\nu(\phi-\phi_0)}\\
-i T e^{i\nu(\phi-\phi_0)}&
C^\ast
\end{pmatrix}
\end{equation}
where

\begin{equation}
C = \frac{1}{2}\left(f_+^2 e^{-iE_{eff}\delta t} + f_-^2 e^{iE_{eff}\delta t}\right)
\end{equation}
and

\begin{equation}
T = f_+ f_- \sin(E_{eff}\delta t)
\end{equation}
As discussed in the main text, the crucial feature of this propagation matrix is the presence in the off-diagonal blocks of the phase $e^{i\nu(\phi-\phi_0)}$ that combines (i) valley discriminating physics through $\nu$, (ii) the polarisation angle of the linear light through $\phi$ and, less importantly for our considerations, (iii) the azimuthal angle of the initial state before the laser. (For dynamics at the K point this is undefined except in the limit of fixed $\phi_0$ and then $k\to 0$.)

The dynamics of two pulses evidently requires multiplication of two of these single pulse propagation matrices with the appropriate unitary evolution of the dynamical phase between the pulses. This can be written as

\begin{equation}
M_T = M(\phi_2)
\begin{pmatrix}
e^{-i\Delta_{gap} \delta T} & 0 \\
0 & e^{i\Delta_{gap} \delta T}
\end{pmatrix}
M(\phi_1)
\end{equation}
and the situation we are interested in is the case of time evolution from the ground state at K to a final excited state

\begin{equation}
\begin{pmatrix} b_v \\ b_c \end{pmatrix} =
M_T \begin{pmatrix} 1 \\ 0 \end{pmatrix}
\end{equation}
Evaluation of the coefficient of the conduction band eigenvector (the square of which obviously gives the final excited charge) after the dynamics is now immediately seen to be

\begin{equation}
c_c = C T e^{-i\Delta_{gap} \delta t}e^{i\nu(\phi_1-\phi_0)} + C^\ast T e^{i\Delta_{gap}\delta t}e^{i\nu(\phi_2-\phi_0)}
\end{equation}
This expression encode the interference of excitation by two pulses with each transition endowing the excited quasiparticles with a phase ($e^{i\nu\phi_{1,2}}$) combining the valley degree of freedom and the polarization angle of the linear light.

Just as for the case of the full calculation shown in the previous section, these can interfere destructively or constructively, again seen by considering the probability for excitation $|c_c|^2$

\begin{equation}
|c_c|^2 = 4|C|^2 T^2 \cos^2(\Delta \delta T - \nu (\phi_2 -\phi_1)/2).
\end{equation}
(note that the irrelevant $\phi_0$ has cancelled).
This is the expression derived in the paper and reveals that the excitation if half a period out of phase at the two valleys $\nu=\pm1$; when one valley is excited (constructive interference) the other valley will be de-excited (destructive interference).

\subsubsection{Rectangular pulse approximation for the multi-cycle pulse}

\begin{figure*}
\includegraphics[width=0.95\textwidth]{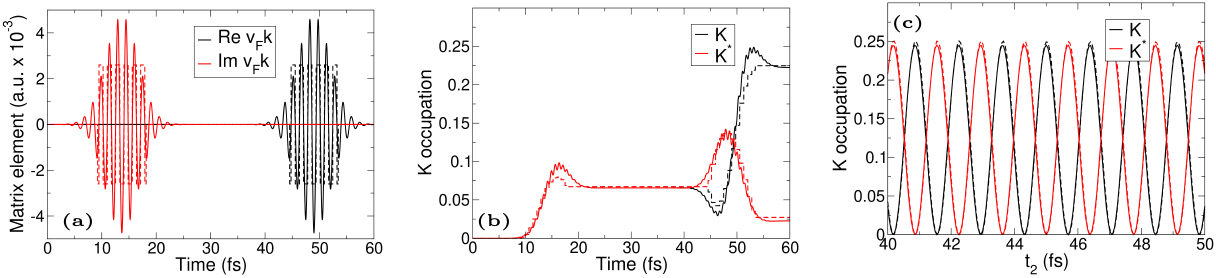}
\caption{Multi-cycle Gaussian pulse with rectangular waveform approximation. Panel (a): Vector potential of the Gaussian envelope multi-cycle pulse and corresponding rectangular waveform pulse. Panel (b): charge excitation at the K and K$^ast$ valleys for the Gaussian pulses (full lines) and rectangular waveform pulses (dashed lines), evidently revealing that these two pulses share similar charge dynamics. Panel (c) Oscillating occupation of the K and K$^\ast$ points as a function of the pulse envelope centre of the second pulse.
}
\label{fig:ms}
\end{figure*}

Having discussed the simplest single cycle pulse, in Fig.~\ref{fig:ms} we now show the case of a multi-cycle pulse. Interestingly, for this more complex case the charge dynamics at the high symmetry K points can also be well reproduce by a rectangular waveform.

\subsection{Time dependent density functional theory methodology}

{\bf Methodology}: Underpinning time-dependent density functional theory is the Runge-Gross theorem \cite{RG84} which establishes that the time-dependent external potential is a unique functional of the time dependent density, given the initial state. Based on this theorem, a system of non-interacting particles can be chosen such that the density of this non-interacting system is equal to that of the interacting system for all times\cite{EFB09,C11,SDG14}, with the wave function of this non-interacting system represented by a Slater determinant of single-particle orbitals. In a fully non-collinear spin-dependent version of these theorems\cite{KDES15} time-dependent Kohn-Sham (KS) orbitals are Pauli spinors governed by the Schr\"odinger equation:

\begin{eqnarray}
i\frac{\partial \psi_j({\bf r},t)}{\partial t}&=&
\Bigg[
\frac{1}{2}\left(-i{\nabla} +\frac{1}{c}{\bf A}_{\rm ext}(t)\right)^2 +v_{s}({\bf r},t) + \frac{1}{2c} {\sigma}\cdot{\bf B}_{s}({\bf r},t) + \nonumber \\
&&\frac{1}{4c^2} {\sigma}\cdot ({\nabla}v_{s}({\bf r},t) \times -i{\nabla})\Bigg]
\psi_j({\bf r},t)
\label{KS}
\end{eqnarray}
where ${\bf A}_{\rm ext}(t)$ is a vector potential representing the applied laser field, and ${\sigma}$ are the Pauli matrices. The KS effective potential $v_{s}({\bf r},t) = v_{\rm ext}({\bf r},t)+v_{\rm H}({\bf r},t)+v_{\rm xc}({\bf r},t)$ is decomposed into the external potential $v_{\rm ext}$, the classical electrostatic Hartree potential $v_{\rm H}$ and the exchange-correlation (XC) potential $v_{\rm xc}$. Similarly, the KS magnetic field is written as ${\bf B}_{s}({\bf r},t)={\bf B}_{\rm ext}(t)+{\bf B}_{\rm xc}({\bf r},t)$ where ${\bf B}_{\rm ext}(t)$ is the magnetic field of the applied laser pulse plus possibly an additional magnetic field and ${\bf B}_{\rm xc}({\bf r},t)$ is the exchange-correlation (XC) magnetic field. The final term of Eq.~\eqref{KS} is the spin-orbit coupling term. It is assumed that the wavelength of the applied laser is much greater than the size of a unit cell and the dipole approximation can be used i.e. the spatial dependence of the vector potential is disregarded. All the implementations are performed using the state-of-the art full potential linearized augmented plane wave (LAPW) method. Within this method the core electrons are treated fully relativistically by solving the radial Dirac equation while higher lying electrons are treated using the scalar relativistic Hamiltonian in the presence of the spin-orbit coupling. 

{\bf Numerical details}: A fully non-collinear version of TDDFT as implemented within the Elk code\cite{elk} is used for all calculations. For these calculations we have used a {\bf k}-point grid of $30\times30\times1$. All states upto 24~eV above the Fermi energy are included and for time propagation we have used a time step of 2.42 atto-seconds, more details of the time propagation algorithm can be found in Ref. \cite{dewhurst_efficient_2016}. A smearing width of 0.027~eV was employed for the ground-state as well as for time propagation. In all calculations we use the adiabatic LDA functional, which has been shown to be highly accurate in treating very early time spin dynamics.

{\bf Method for calculation of $N_{ex}$}: To calculate the crystal momentum, $\v k$-resolved excitation, $N_{\text{ex}}$, we use the expression

\begin{equation}
\label{nex}
N_{\text{ex}}({\v k}) = \sum_a^{occ}\sum_b^{unocc} \left|\braket{\psi_{a{\v k}}(t)|\psi_{b{\v k}}(t=0)}\right|^2,
\end{equation}
where $a$, $b$ are the band indices. The time-dependent KS orbitals, $\psi_{a\v k}(t)$, at a given time $t$ are projected on to the ground-state orbitals, $\psi_{b\v k}(t=0)$, to calculate the change in occupation of the KS orbitals. Formally, within TDDFT, the transient occupation of the excited-states does not necessarily follow that of the KS system. For weakly excited systems, however, the difference is expected to be small. It should be noted that for high fluence pulses where the band renormalization effects are large, not applicable to this work, such an approximation would fail.

\subsection{Dependence of valley polarization of pulse amplitude and frequency}

\begin{figure*}
\includegraphics[width=0.95\textwidth]{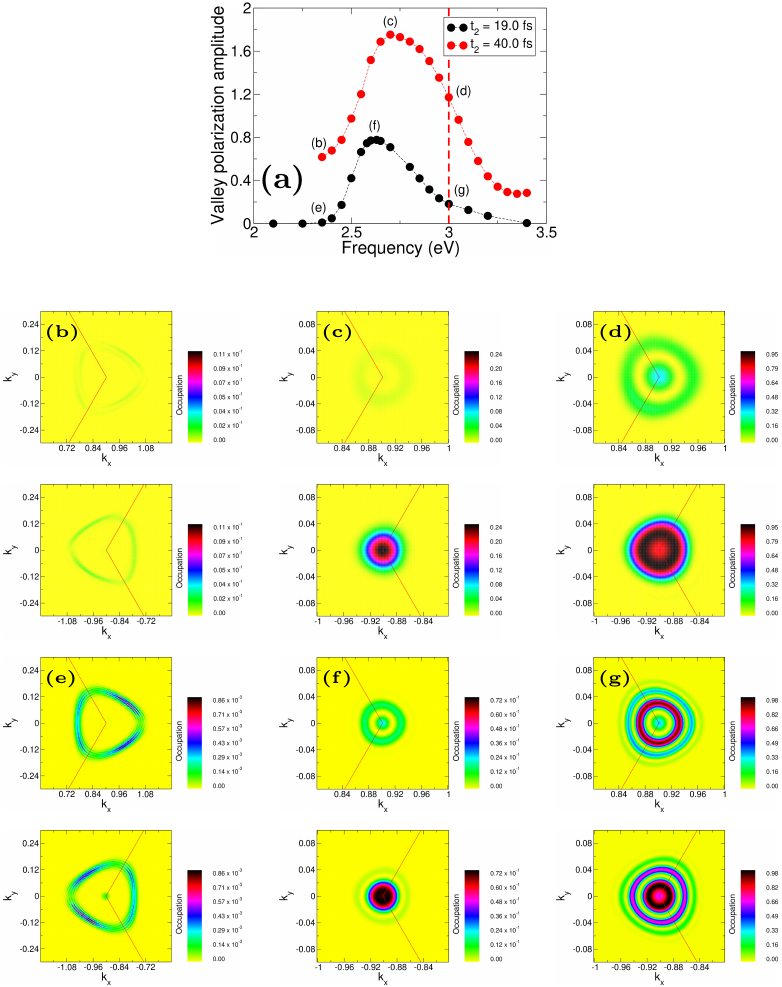}
\caption{Dependence of valley polarisation oscillation amplitude on frequency.
}
\label{F}
\end{figure*}

\begin{figure*}
\includegraphics[width=0.95\textwidth]{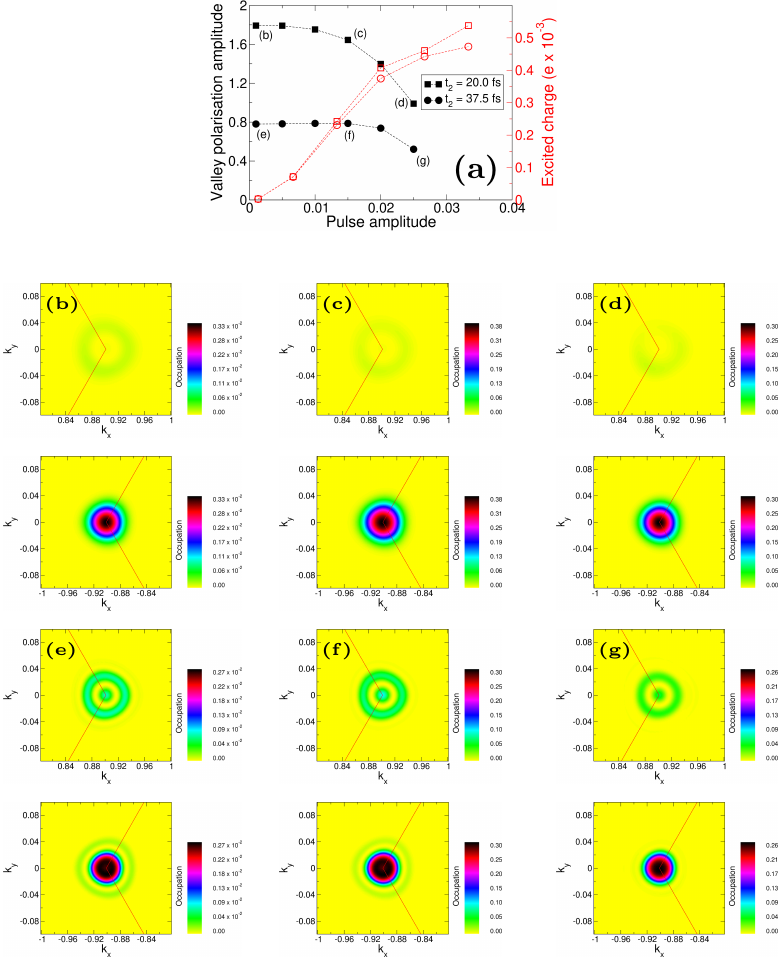}
\caption{Dependence on polarisation oscillation amplitude on amplitude.
}
\label{A}
\end{figure*}

Here we show additional details of the dependence of the amplitude of the oscillation of valley polarisation on pulse frequency, see Fig.~\ref{F}, and pulse amplitude, see Fig.~\ref{A}. In the first of these figures panel (a) shows the amplitude of the oscillation for two cases of the full width half maximum 7~fs pulse shown in panel (a) of Fig.~1 of the main text, with the time of the second pulse taken to be 19~fs and 40~fs respectively. Shown below are $\v k$-resolved charge excitation at the K and K$^\ast$ valleys for a series of points taken to represent the maximum valley polarisation at three frequencies corresponding to cases of strong valley oscillation amplitude, and weak valley oscillation amplitude. Similar data is then presented for the case of the dependence on pulse amplitude, see Fig.~\ref{A}. However in panel (a) of this figure is shown also the excited charge  as well as the valley polarisation.

%


\begin{thebibliography}{24}%
\makeatletter
\providecommand \@ifxundefined [1]{%
 \@ifx{#1\undefined}
}%
\providecommand \@ifnum [1]{%
 \ifnum #1\expandafter \@firstoftwo
 \else \expandafter \@secondoftwo
 \fi
}%
\providecommand \@ifx [1]{%
 \ifx #1\expandafter \@firstoftwo
 \else \expandafter \@secondoftwo
 \fi
}%
\providecommand \natexlab [1]{#1}%
\providecommand \enquote  [1]{``#1''}%
\providecommand \bibnamefont  [1]{#1}%
\providecommand \bibfnamefont [1]{#1}%
\providecommand \citenamefont [1]{#1}%
\providecommand \href@noop [0]{\@secondoftwo}%
\providecommand \href [0]{\begingroup \@sanitize@url \@href}%
\providecommand \@href[1]{\@@startlink{#1}\@@href}%
\providecommand \@@href[1]{\endgroup#1\@@endlink}%
\providecommand \@sanitize@url [0]{\catcode `\\12\catcode `\$12\catcode
  `\&12\catcode `\#12\catcode `\^12\catcode `\_12\catcode `\%12\relax}%
\providecommand \@@startlink[1]{}%
\providecommand \@@endlink[0]{}%
\providecommand \url  [0]{\begingroup\@sanitize@url \@url }%
\providecommand \@url [1]{\endgroup\@href {#1}{\urlprefix }}%
\providecommand \urlprefix  [0]{URL }%
\providecommand \Eprint [0]{\href }%
\providecommand \doibase [0]{http://dx.doi.org/}%
\providecommand \selectlanguage [0]{\@gobble}%
\providecommand \bibinfo  [0]{\@secondoftwo}%
\providecommand \bibfield  [0]{\@secondoftwo}%
\providecommand \translation [1]{[#1]}%
\providecommand \BibitemOpen [0]{}%
\providecommand \bibitemStop [0]{}%
\providecommand \bibitemNoStop [0]{.\EOS\space}%
\providecommand \EOS [0]{\spacefactor3000\relax}%
\providecommand \BibitemShut  [1]{\csname bibitem#1\endcsname}%
\let\auto@bib@innerbib\@empty
\bibitem [{\citenamefont {Suzuki}\ \emph {et~al.}(2014)\citenamefont {Suzuki},
  \citenamefont {Sakano}, \citenamefont {Zhang}, \citenamefont {Akashi},
  \citenamefont {Morikawa}, \citenamefont {Harasawa}, \citenamefont {Yaji},
  \citenamefont {Kuroda}, \citenamefont {Miyamoto}, \citenamefont {Okuda},
  \citenamefont {Ishizaka}, \citenamefont {Arita},\ and\ \citenamefont
  {Iwasa}}]{SSZA14}%
  \BibitemOpen
  \bibfield  {author} {\bibinfo {author} {\bibfnamefont {R.}~\bibnamefont
  {Suzuki}}, \bibinfo {author} {\bibfnamefont {M.}~\bibnamefont {Sakano}},
  \bibinfo {author} {\bibfnamefont {Y.~J.}\ \bibnamefont {Zhang}}, \bibinfo
  {author} {\bibfnamefont {R.}~\bibnamefont {Akashi}}, \bibinfo {author}
  {\bibfnamefont {D.}~\bibnamefont {Morikawa}}, \bibinfo {author}
  {\bibfnamefont {A.}~\bibnamefont {Harasawa}}, \bibinfo {author}
  {\bibfnamefont {K.}~\bibnamefont {Yaji}}, \bibinfo {author} {\bibfnamefont
  {K.}~\bibnamefont {Kuroda}}, \bibinfo {author} {\bibfnamefont
  {K.}~\bibnamefont {Miyamoto}}, \bibinfo {author} {\bibfnamefont
  {T.}~\bibnamefont {Okuda}}, \bibinfo {author} {\bibfnamefont
  {K.}~\bibnamefont {Ishizaka}}, \bibinfo {author} {\bibfnamefont
  {R.}~\bibnamefont {Arita}}, \ and\ \bibinfo {author} {\bibfnamefont
  {Y.}~\bibnamefont {Iwasa}},\ }\href {\doibase 10.1038/nnano.2014.148}
  {\bibfield  {journal} {\bibinfo  {journal} {Nature Nanotechnology}\ }\textbf
  {\bibinfo {volume} {9}},\ \bibinfo {pages} {611} (\bibinfo {year}
  {2014})}\BibitemShut {NoStop}%
\bibitem [{\citenamefont {Mak}\ \emph {et~al.}(2018)\citenamefont {Mak},
  \citenamefont {Xiao},\ and\ \citenamefont {Shan}}]{mak_lightvalley_2018}%
  \BibitemOpen
  \bibfield  {author} {\bibinfo {author} {\bibfnamefont {K.~F.}\ \bibnamefont
  {Mak}}, \bibinfo {author} {\bibfnamefont {D.}~\bibnamefont {Xiao}}, \ and\
  \bibinfo {author} {\bibfnamefont {J.}~\bibnamefont {Shan}},\ }\href {\doibase
  10.1038/s41566-018-0204-6} {\bibfield  {journal} {\bibinfo  {journal} {Nature
  Photonics}\ }\textbf {\bibinfo {volume} {12}},\ \bibinfo {pages} {451}
  (\bibinfo {year} {2018})},\ \bibinfo {note} {number: 8 Publisher: Nature
  Publishing Group}\BibitemShut {NoStop}%
\bibitem [{\citenamefont {Chen}\ \emph {et~al.}(2020)\citenamefont {Chen},
  \citenamefont {Lo}, \citenamefont {Fan}, \citenamefont {Wang}, \citenamefont
  {Huang},\ and\ \citenamefont {Lei}}]{chen_chiral_2020}%
  \BibitemOpen
  \bibfield  {author} {\bibinfo {author} {\bibfnamefont {P.}~\bibnamefont
  {Chen}}, \bibinfo {author} {\bibfnamefont {T.~W.}\ \bibnamefont {Lo}},
  \bibinfo {author} {\bibfnamefont {Y.}~\bibnamefont {Fan}}, \bibinfo {author}
  {\bibfnamefont {S.}~\bibnamefont {Wang}}, \bibinfo {author} {\bibfnamefont
  {H.}~\bibnamefont {Huang}}, \ and\ \bibinfo {author} {\bibfnamefont
  {D.}~\bibnamefont {Lei}},\ }\href {\doibase 10.1002/adom.201901233}
  {\bibfield  {journal} {\bibinfo  {journal} {Advanced Optical Materials}\
  }\textbf {\bibinfo {volume} {8}},\ \bibinfo {pages} {1901233} (\bibinfo
  {year} {2020})},\ \bibinfo {note} {\_eprint:
  https://onlinelibrary.wiley.com/doi/pdf/10.1002/adom.201901233}\BibitemShut
  {NoStop}%
\bibitem [{\citenamefont {Settnes}\ \emph {et~al.}(2016)\citenamefont
  {Settnes}, \citenamefont {Power}, \citenamefont {Brandbyge},\ and\
  \citenamefont {Jauho}}]{settnes_graphene_2016}%
  \BibitemOpen
  \bibfield  {author} {\bibinfo {author} {\bibfnamefont {M.}~\bibnamefont
  {Settnes}}, \bibinfo {author} {\bibfnamefont {S.~R.}\ \bibnamefont {Power}},
  \bibinfo {author} {\bibfnamefont {M.}~\bibnamefont {Brandbyge}}, \ and\
  \bibinfo {author} {\bibfnamefont {A.-P.}\ \bibnamefont {Jauho}},\ }\href
  {\doibase 10.1103/PhysRevLett.117.276801} {\bibfield  {journal} {\bibinfo
  {journal} {Phys. Rev. Lett.}\ }\textbf {\bibinfo {volume} {117}},\ \bibinfo
  {pages} {276801} (\bibinfo {year} {2016})}\BibitemShut {NoStop}%
\bibitem [{\citenamefont {Gupta}\ \emph {et~al.}(2019)\citenamefont {Gupta},
  \citenamefont {Rost}, \citenamefont {Fleischmann}, \citenamefont {Sharma},\
  and\ \citenamefont {Shallcross}}]{gupta_straintronics_2019}%
  \BibitemOpen
  \bibfield  {author} {\bibinfo {author} {\bibfnamefont {R.}~\bibnamefont
  {Gupta}}, \bibinfo {author} {\bibfnamefont {F.}~\bibnamefont {Rost}},
  \bibinfo {author} {\bibfnamefont {M.}~\bibnamefont {Fleischmann}}, \bibinfo
  {author} {\bibfnamefont {S.}~\bibnamefont {Sharma}}, \ and\ \bibinfo {author}
  {\bibfnamefont {S.}~\bibnamefont {Shallcross}},\ }\href {\doibase
  10.1103/PhysRevB.99.125407} {\bibfield  {journal} {\bibinfo  {journal}
  {Physical Review B}\ }\textbf {\bibinfo {volume} {99}},\ \bibinfo {pages}
  {125407} (\bibinfo {year} {2019})}\BibitemShut {NoStop}%
\bibitem [{\citenamefont {Zhao}\ \emph {et~al.}(2022)\citenamefont {Zhao},
  \citenamefont {Wang}, \citenamefont {Wang}, \citenamefont {Zhou},
  \citenamefont {Zhang}, \citenamefont {Cui}, \citenamefont {Wang},
  \citenamefont {Liu}, \citenamefont {Han}, \citenamefont {Luo}, \citenamefont
  {Yue}, \citenamefont {Song},\ and\ \citenamefont
  {Sun}}]{zhao_ultrafast_2022}%
  \BibitemOpen
  \bibfield  {author} {\bibinfo {author} {\bibfnamefont {L.-Y.}\ \bibnamefont
  {Zhao}}, \bibinfo {author} {\bibfnamefont {H.}~\bibnamefont {Wang}}, \bibinfo
  {author} {\bibfnamefont {H.-Y.}\ \bibnamefont {Wang}}, \bibinfo {author}
  {\bibfnamefont {Q.}~\bibnamefont {Zhou}}, \bibinfo {author} {\bibfnamefont
  {X.-L.}\ \bibnamefont {Zhang}}, \bibinfo {author} {\bibfnamefont
  {T.}~\bibnamefont {Cui}}, \bibinfo {author} {\bibfnamefont {L.}~\bibnamefont
  {Wang}}, \bibinfo {author} {\bibfnamefont {T.-Y.}\ \bibnamefont {Liu}},
  \bibinfo {author} {\bibfnamefont {Y.-X.}\ \bibnamefont {Han}}, \bibinfo
  {author} {\bibfnamefont {Y.}~\bibnamefont {Luo}}, \bibinfo {author}
  {\bibfnamefont {Y.-Y.}\ \bibnamefont {Yue}}, \bibinfo {author} {\bibfnamefont
  {M.-S.}\ \bibnamefont {Song}}, \ and\ \bibinfo {author} {\bibfnamefont
  {H.-B.}\ \bibnamefont {Sun}},\ }\href {\doibase 10.1186/s43074-022-00049-1}
  {\bibfield  {journal} {\bibinfo  {journal} {PhotoniX}\ }\textbf {\bibinfo
  {volume} {3}},\ \bibinfo {pages} {5} (\bibinfo {year} {2022})}\BibitemShut
  {NoStop}%
\bibitem [{\citenamefont {Yang}\ \emph {et~al.}(2019)\citenamefont {Yang},
  \citenamefont {Aghaeimeibodi},\ and\ \citenamefont
  {Waks}}]{yang_chiral_2019}%
  \BibitemOpen
  \bibfield  {author} {\bibinfo {author} {\bibfnamefont {Z.}~\bibnamefont
  {Yang}}, \bibinfo {author} {\bibfnamefont {S.}~\bibnamefont {Aghaeimeibodi}},
  \ and\ \bibinfo {author} {\bibfnamefont {E.}~\bibnamefont {Waks}},\ }\href
  {\doibase 10.1364/OE.27.021367} {\bibfield  {journal} {\bibinfo  {journal}
  {Optics Express}\ }\textbf {\bibinfo {volume} {27}},\ \bibinfo {pages}
  {21367} (\bibinfo {year} {2019})},\ \bibinfo {note} {publisher: Optica
  Publishing Group}\BibitemShut {NoStop}%
\bibitem [{\citenamefont {Xiao}\ \emph {et~al.}(2012)\citenamefont {Xiao},
  \citenamefont {Liu}, \citenamefont {Feng}, \citenamefont {Xu},\ and\
  \citenamefont {Yao}}]{xiao_coupled_2012}%
  \BibitemOpen
  \bibfield  {author} {\bibinfo {author} {\bibfnamefont {D.}~\bibnamefont
  {Xiao}}, \bibinfo {author} {\bibfnamefont {G.-B.}\ \bibnamefont {Liu}},
  \bibinfo {author} {\bibfnamefont {W.}~\bibnamefont {Feng}}, \bibinfo {author}
  {\bibfnamefont {X.}~\bibnamefont {Xu}}, \ and\ \bibinfo {author}
  {\bibfnamefont {W.}~\bibnamefont {Yao}},\ }\href {\doibase
  10.1103/PhysRevLett.108.196802} {\bibfield  {journal} {\bibinfo  {journal}
  {Physical Review Letters}\ }\textbf {\bibinfo {volume} {108}},\ \bibinfo
  {pages} {196802} (\bibinfo {year} {2012})}\BibitemShut {NoStop}%
\bibitem [{\citenamefont {Mak}\ \emph {et~al.}(2012)\citenamefont {Mak},
  \citenamefont {He}, \citenamefont {Shan},\ and\ \citenamefont
  {Heinz}}]{mak_control_2012}%
  \BibitemOpen
  \bibfield  {author} {\bibinfo {author} {\bibfnamefont {K.~F.}\ \bibnamefont
  {Mak}}, \bibinfo {author} {\bibfnamefont {K.}~\bibnamefont {He}}, \bibinfo
  {author} {\bibfnamefont {J.}~\bibnamefont {Shan}}, \ and\ \bibinfo {author}
  {\bibfnamefont {T.~F.}\ \bibnamefont {Heinz}},\ }\href {\doibase
  10.1038/nnano.2012.96} {\bibfield  {journal} {\bibinfo  {journal} {Nature
  Nanotechnology}\ }\textbf {\bibinfo {volume} {7}},\ \bibinfo {pages} {494}
  (\bibinfo {year} {2012})}\BibitemShut {NoStop}%
\bibitem [{\citenamefont {Xiao}\ \emph {et~al.}(2015)\citenamefont {Xiao},
  \citenamefont {Ye}, \citenamefont {Wang}, \citenamefont {Zhu}, \citenamefont
  {Wang},\ and\ \citenamefont {Zhang}}]{xiao_nonlinear_2015}%
  \BibitemOpen
  \bibfield  {author} {\bibinfo {author} {\bibfnamefont {J.}~\bibnamefont
  {Xiao}}, \bibinfo {author} {\bibfnamefont {Z.}~\bibnamefont {Ye}}, \bibinfo
  {author} {\bibfnamefont {Y.}~\bibnamefont {Wang}}, \bibinfo {author}
  {\bibfnamefont {H.}~\bibnamefont {Zhu}}, \bibinfo {author} {\bibfnamefont
  {Y.}~\bibnamefont {Wang}}, \ and\ \bibinfo {author} {\bibfnamefont
  {X.}~\bibnamefont {Zhang}},\ }\href {\doibase 10.1038/lsa.2015.139}
  {\bibfield  {journal} {\bibinfo  {journal} {Light: Science \& Applications}\
  }\textbf {\bibinfo {volume} {4}},\ \bibinfo {pages} {e366} (\bibinfo {year}
  {2015})}\BibitemShut {NoStop}%
\bibitem [{\citenamefont {Langer}\ \emph {et~al.}(2018)\citenamefont {Langer},
  \citenamefont {Schmid}, \citenamefont {Schlauderer}, \citenamefont {Gmitra},
  \citenamefont {Fabian}, \citenamefont {Nagler}, \citenamefont {Schüller},
  \citenamefont {Korn}, \citenamefont {Hawkins}, \citenamefont {Steiner},
  \citenamefont {Huttner}, \citenamefont {Koch}, \citenamefont {Kira},\ and\
  \citenamefont {Huber}}]{langer_lightwave_2018}%
  \BibitemOpen
  \bibfield  {author} {\bibinfo {author} {\bibfnamefont {F.}~\bibnamefont
  {Langer}}, \bibinfo {author} {\bibfnamefont {C.~P.}\ \bibnamefont {Schmid}},
  \bibinfo {author} {\bibfnamefont {S.}~\bibnamefont {Schlauderer}}, \bibinfo
  {author} {\bibfnamefont {M.}~\bibnamefont {Gmitra}}, \bibinfo {author}
  {\bibfnamefont {J.}~\bibnamefont {Fabian}}, \bibinfo {author} {\bibfnamefont
  {P.}~\bibnamefont {Nagler}}, \bibinfo {author} {\bibfnamefont
  {C.}~\bibnamefont {Schüller}}, \bibinfo {author} {\bibfnamefont
  {T.}~\bibnamefont {Korn}}, \bibinfo {author} {\bibfnamefont {P.~G.}\
  \bibnamefont {Hawkins}}, \bibinfo {author} {\bibfnamefont {J.~T.}\
  \bibnamefont {Steiner}}, \bibinfo {author} {\bibfnamefont {U.}~\bibnamefont
  {Huttner}}, \bibinfo {author} {\bibfnamefont {S.~W.}\ \bibnamefont {Koch}},
  \bibinfo {author} {\bibfnamefont {M.}~\bibnamefont {Kira}}, \ and\ \bibinfo
  {author} {\bibfnamefont {R.}~\bibnamefont {Huber}},\ }\href {\doibase
  10.1038/s41586-018-0013-6} {\bibfield  {journal} {\bibinfo  {journal}
  {Nature}\ }\textbf {\bibinfo {volume} {557}},\ \bibinfo {pages} {76}
  (\bibinfo {year} {2018})}\BibitemShut {NoStop}%
\bibitem [{\citenamefont {Bergh\"auser}\ \emph {et~al.}(2018)\citenamefont
  {Bergh\"auser}, \citenamefont {Bernal-Villamil}, \citenamefont {Schmidt},
  \citenamefont {Schneider}, \citenamefont {Niehues}, \citenamefont {Erhart},
  \citenamefont {Michaelis~de Vasconcellos}, \citenamefont {Bratschitsch},
  \citenamefont {Knorr},\ and\ \citenamefont
  {Malic}}]{berghauser_inverted_2018}%
  \BibitemOpen
  \bibfield  {author} {\bibinfo {author} {\bibfnamefont {G.}~\bibnamefont
  {Bergh\"auser}}, \bibinfo {author} {\bibfnamefont {I.}~\bibnamefont
  {Bernal-Villamil}}, \bibinfo {author} {\bibfnamefont {R.}~\bibnamefont
  {Schmidt}}, \bibinfo {author} {\bibfnamefont {R.}~\bibnamefont {Schneider}},
  \bibinfo {author} {\bibfnamefont {I.}~\bibnamefont {Niehues}}, \bibinfo
  {author} {\bibfnamefont {P.}~\bibnamefont {Erhart}}, \bibinfo {author}
  {\bibfnamefont {S.}~\bibnamefont {Michaelis~de Vasconcellos}}, \bibinfo
  {author} {\bibfnamefont {R.}~\bibnamefont {Bratschitsch}}, \bibinfo {author}
  {\bibfnamefont {A.}~\bibnamefont {Knorr}}, \ and\ \bibinfo {author}
  {\bibfnamefont {E.}~\bibnamefont {Malic}},\ }\href {\doibase
  10.1038/s41467-018-03354-1} {\bibfield  {journal} {\bibinfo  {journal}
  {Nature Communications}\ }\textbf {\bibinfo {volume} {9}},\ \bibinfo {pages}
  {971} (\bibinfo {year} {2018})},\ \bibinfo {note} {number: 1 Publisher:
  Nature Publishing Group}\BibitemShut {NoStop}%
\bibitem [{\citenamefont {Ishii}\ \emph {et~al.}(2019)\citenamefont {Ishii},
  \citenamefont {Yokoshi},\ and\ \citenamefont
  {Ishihara}}]{ishii_optical_2019}%
  \BibitemOpen
  \bibfield  {author} {\bibinfo {author} {\bibfnamefont {S.}~\bibnamefont
  {Ishii}}, \bibinfo {author} {\bibfnamefont {N.}~\bibnamefont {Yokoshi}}, \
  and\ \bibinfo {author} {\bibfnamefont {H.}~\bibnamefont {Ishihara}},\ }\href
  {\doibase 10.1088/1742-6596/1220/1/012056} {\bibfield  {journal} {\bibinfo
  {journal} {Journal of Physics: Conference Series}\ }\textbf {\bibinfo
  {volume} {1220}},\ \bibinfo {pages} {012056} (\bibinfo {year} {2019})},\
  \bibinfo {note} {publisher: IOP Publishing}\BibitemShut {NoStop}%
\bibitem [{\citenamefont {Cho}\ \emph {et~al.}(2018)\citenamefont {Cho},
  \citenamefont {Park}, \citenamefont {Hong}, \citenamefont {Jung},
  \citenamefont {Kim}, \citenamefont {Han}, \citenamefont {Kyung},
  \citenamefont {Kim}, \citenamefont {Mo}, \citenamefont {Denlinger},
  \citenamefont {Shim}, \citenamefont {Han}, \citenamefont {Kim},\ and\
  \citenamefont {Park}}]{cho_experimental_2018}%
  \BibitemOpen
  \bibfield  {author} {\bibinfo {author} {\bibfnamefont {S.}~\bibnamefont
  {Cho}}, \bibinfo {author} {\bibfnamefont {J.-H.}\ \bibnamefont {Park}},
  \bibinfo {author} {\bibfnamefont {J.}~\bibnamefont {Hong}}, \bibinfo {author}
  {\bibfnamefont {J.}~\bibnamefont {Jung}}, \bibinfo {author} {\bibfnamefont
  {B.~S.}\ \bibnamefont {Kim}}, \bibinfo {author} {\bibfnamefont
  {G.}~\bibnamefont {Han}}, \bibinfo {author} {\bibfnamefont {W.}~\bibnamefont
  {Kyung}}, \bibinfo {author} {\bibfnamefont {Y.}~\bibnamefont {Kim}}, \bibinfo
  {author} {\bibfnamefont {S.-K.}\ \bibnamefont {Mo}}, \bibinfo {author}
  {\bibfnamefont {J.}~\bibnamefont {Denlinger}}, \bibinfo {author}
  {\bibfnamefont {J.~H.}\ \bibnamefont {Shim}}, \bibinfo {author}
  {\bibfnamefont {J.~H.}\ \bibnamefont {Han}}, \bibinfo {author} {\bibfnamefont
  {C.}~\bibnamefont {Kim}}, \ and\ \bibinfo {author} {\bibfnamefont {S.~R.}\
  \bibnamefont {Park}},\ }\href {\doibase 10.1103/PhysRevLett.121.186401}
  {\bibfield  {journal} {\bibinfo  {journal} {Physical Review Letters}\
  }\textbf {\bibinfo {volume} {121}},\ \bibinfo {pages} {186401} (\bibinfo
  {year} {2018})},\ \bibinfo {note} {publisher: American Physical
  Society}\BibitemShut {NoStop}%
\bibitem [{\citenamefont {Motlagh}\ and\ \citenamefont
  {Apalkov}(2021)}]{MotlaghApalkov}%
  \BibitemOpen
  \bibfield  {author} {\bibinfo {author} {\bibfnamefont {S.~A.~O.}\
  \bibnamefont {Motlagh}}\ and\ \bibinfo {author} {\bibfnamefont
  {V.}~\bibnamefont {Apalkov}},\ }\href {\doibase doi:10.1515/nanoph-2021-0227}
  {\bibfield  {journal} {\bibinfo  {journal} {Nanophotonics}\ }\textbf
  {\bibinfo {volume} {10}},\ \bibinfo {pages} {3677} (\bibinfo {year}
  {2021})}\BibitemShut {NoStop}%
\bibitem [{\citenamefont {Hanbicki}\ \emph {et~al.}(2016)\citenamefont
  {Hanbicki}, \citenamefont {McCreary}, \citenamefont {Kioseoglou},
  \citenamefont {Currie}, \citenamefont {Hellberg}, \citenamefont {Friedman},\
  and\ \citenamefont {Jonker}}]{hanbicki_high_2016}%
  \BibitemOpen
  \bibfield  {author} {\bibinfo {author} {\bibfnamefont {A.~T.}\ \bibnamefont
  {Hanbicki}}, \bibinfo {author} {\bibfnamefont {K.~M.}\ \bibnamefont
  {McCreary}}, \bibinfo {author} {\bibfnamefont {G.}~\bibnamefont
  {Kioseoglou}}, \bibinfo {author} {\bibfnamefont {M.}~\bibnamefont {Currie}},
  \bibinfo {author} {\bibfnamefont {C.~S.}\ \bibnamefont {Hellberg}}, \bibinfo
  {author} {\bibfnamefont {A.~L.}\ \bibnamefont {Friedman}}, \ and\ \bibinfo
  {author} {\bibfnamefont {B.~T.}\ \bibnamefont {Jonker}},\ }\href {\doibase
  10.1063/1.4942797} {\bibfield  {journal} {\bibinfo  {journal} {AIP Advances}\
  }\textbf {\bibinfo {volume} {6}},\ \bibinfo {pages} {055804} (\bibinfo {year}
  {2016})},\ \bibinfo {note} {publisher: American Institute of
  Physics}\BibitemShut {NoStop}%
\bibitem [{\citenamefont {Pashalou}\ and\ \citenamefont
  {Goudarzi}(2020)}]{pashalou_coherent_2020}%
  \BibitemOpen
  \bibfield  {author} {\bibinfo {author} {\bibfnamefont {S.}~\bibnamefont
  {Pashalou}}\ and\ \bibinfo {author} {\bibfnamefont {H.}~\bibnamefont
  {Goudarzi}},\ }\href {\doibase 10.1016/j.spmi.2020.106566} {\bibfield
  {journal} {\bibinfo  {journal} {Superlattices and Microstructures}\ }\textbf
  {\bibinfo {volume} {144}},\ \bibinfo {pages} {106566} (\bibinfo {year}
  {2020})}\BibitemShut {NoStop}%
\bibitem [{\citenamefont {Runge}\ and\ \citenamefont {Gross}(1984)}]{RG84}%
  \BibitemOpen
  \bibfield  {author} {\bibinfo {author} {\bibfnamefont {E.}~\bibnamefont
  {Runge}}\ and\ \bibinfo {author} {\bibfnamefont {E.}~\bibnamefont {Gross}},\
  }\href@noop {} {\bibfield  {journal} {\bibinfo  {journal} {Phys. Rev. Lett.}\
  }\textbf {\bibinfo {volume} {52}},\ \bibinfo {pages} {997} (\bibinfo {year}
  {1984})}\BibitemShut {NoStop}%
\bibitem [{\citenamefont {Elliott}\ \emph {et~al.}(2009)\citenamefont
  {Elliott}, \citenamefont {Furche},\ and\ \citenamefont {Burke}}]{EFB09}%
  \BibitemOpen
  \bibfield  {author} {\bibinfo {author} {\bibfnamefont {P.}~\bibnamefont
  {Elliott}}, \bibinfo {author} {\bibfnamefont {F.}~\bibnamefont {Furche}}, \
  and\ \bibinfo {author} {\bibfnamefont {K.}~\bibnamefont {Burke}},\ }in\
  \href@noop {} {\emph {\bibinfo {booktitle} {Reviews in Computational
  Chemistry}}},\ Vol.~\bibinfo {volume} {26},\ \bibinfo {editor} {edited by\
  \bibinfo {editor} {\bibfnamefont {K.}~\bibnamefont {Lipkowitz}}\ and\
  \bibinfo {editor} {\bibfnamefont {T.}~\bibnamefont {Cundari}}}\ (\bibinfo
  {publisher} {Wiley, Hoboken, NJ},\ \bibinfo {year} {2009})\ pp.\ \bibinfo
  {pages} {91--165}\BibitemShut {NoStop}%
\bibitem [{\citenamefont {Ullrich}(2011)}]{C11}%
  \BibitemOpen
  \bibfield  {author} {\bibinfo {author} {\bibfnamefont {C.~A.}\ \bibnamefont
  {Ullrich}},\ }\href@noop {} {\emph {\bibinfo {title} {Time-Dependent
  Density-Functional Theory Concepts and Applications}}}\ (\bibinfo
  {publisher} {Oxford University Press},\ \bibinfo {address} {Oxford, New
  York},\ \bibinfo {year} {2011})\BibitemShut {NoStop}%
\bibitem [{\citenamefont {Sharma}\ \emph {et~al.}(2014)\citenamefont {Sharma},
  \citenamefont {Dewhurst},\ and\ \citenamefont {Gross}}]{SDG14}%
  \BibitemOpen
  \bibfield  {author} {\bibinfo {author} {\bibfnamefont {S.}~\bibnamefont
  {Sharma}}, \bibinfo {author} {\bibfnamefont {J.~K.}\ \bibnamefont
  {Dewhurst}}, \ and\ \bibinfo {author} {\bibfnamefont {E.~K.~U.}\ \bibnamefont
  {Gross}},\ }in\ \href {\doibase 10.1007/128_2014_529} {\emph {\bibinfo
  {booktitle} {First Principles Approaches to Spectroscopic Properties of
  Complex Materials}}},\ \bibinfo {editor} {edited by\ \bibinfo {editor}
  {\bibfnamefont {C.}~\bibnamefont {Di~Valentin}}, \bibinfo {editor}
  {\bibfnamefont {S.}~\bibnamefont {Botti}}, \ and\ \bibinfo {editor}
  {\bibfnamefont {M.}~\bibnamefont {Cococcioni}}}\ (\bibinfo  {publisher}
  {Springer Berlin Heidelberg},\ \bibinfo {address} {Berlin, Heidelberg},\
  \bibinfo {year} {2014})\ pp.\ \bibinfo {pages} {235--257}\BibitemShut
  {NoStop}%
\bibitem [{\citenamefont {Krieger}\ \emph {et~al.}(2015)\citenamefont
  {Krieger}, \citenamefont {Dewhurst}, \citenamefont {Elliott}, \citenamefont
  {Sharma},\ and\ \citenamefont {Gross}}]{KDES15}%
  \BibitemOpen
  \bibfield  {author} {\bibinfo {author} {\bibfnamefont {K.}~\bibnamefont
  {Krieger}}, \bibinfo {author} {\bibfnamefont {J.~K.}\ \bibnamefont
  {Dewhurst}}, \bibinfo {author} {\bibfnamefont {P.}~\bibnamefont {Elliott}},
  \bibinfo {author} {\bibfnamefont {S.}~\bibnamefont {Sharma}}, \ and\ \bibinfo
  {author} {\bibfnamefont {E.~K.~U.}\ \bibnamefont {Gross}},\ }\href {\doibase
  10.1021/acs.jctc.5b00621} {\bibfield  {journal} {\bibinfo  {journal} {J.
  Chem. Theory Comput.}\ }\textbf {\bibinfo {volume} {11}},\ \bibinfo {pages}
  {4870} (\bibinfo {year} {2015})}\BibitemShut {NoStop}%
\bibitem [{\citenamefont {Dewhurst}\ \emph {et~al.}(2018)\citenamefont
  {Dewhurst}, \citenamefont {Sharma},\ and\ \citenamefont {et~al.}}]{elk}%
  \BibitemOpen
  \bibfield  {author} {\bibinfo {author} {\bibfnamefont {J.~K.}\ \bibnamefont
  {Dewhurst}}, \bibinfo {author} {\bibfnamefont {S.}~\bibnamefont {Sharma}}, \
  and\ \bibinfo {author} {\bibnamefont {et~al.}},\ }\href {elk.sourceforge.net}
  {} (\bibinfo {year} {Jan. 14 {\bf 2018}})\BibitemShut {NoStop}%
\bibitem [{\citenamefont {Dewhurst}\ \emph {et~al.}(2016)\citenamefont
  {Dewhurst}, \citenamefont {Krieger}, \citenamefont {Sharma},\ and\
  \citenamefont {Gross}}]{dewhurst_efficient_2016}%
  \BibitemOpen
  \bibfield  {author} {\bibinfo {author} {\bibfnamefont {J.~K.}\ \bibnamefont
  {Dewhurst}}, \bibinfo {author} {\bibfnamefont {K.}~\bibnamefont {Krieger}},
  \bibinfo {author} {\bibfnamefont {S.}~\bibnamefont {Sharma}}, \ and\ \bibinfo
  {author} {\bibfnamefont {E.~K.~U.}\ \bibnamefont {Gross}},\ }\href {\doibase
  10.1016/j.cpc.2016.09.001} {\bibfield  {journal} {\bibinfo  {journal}
  {Computer Physics Communications}\ }\textbf {\bibinfo {volume} {209}},\
  \bibinfo {pages} {92} (\bibinfo {year} {2016})}\BibitemShut {NoStop}%
\end{thebibliography}

\end{document}